\documentclass{pasj00}
\usepackage{color}
\color{black}
\tenpoint
\SetRunningHead{K. Sato et al.}{
Suzaku observations of Abell 1246 cluster}

\title{
Temperature and Entropy Profiles to the Virial Radius of
Abell~1246 Cluster Observed with Suzaku
}
\author{
 Kosuke \textsc{Sato},\altaffilmark{1}
 Kyoko \textsc{Matsushita},\altaffilmark{1}
 Noriko \textsc{Y.~Yamasaki},\altaffilmark{2}\\
 Shin \textsc{Sasaki},\altaffilmark{3}
and  Takaya \textsc{Ohashi}\altaffilmark{3}
}
\altaffiltext{1}{
Department of Physics, Tokyo University of Science,
1-3 Kagurazaka, Shinjuku-ku, Tokyo 162-8601}
\email{ksato@rs.tus.ac.jp}
\altaffiltext{2}{
Institute of Space and Astronautical Science (ISAS),
Japan Aerospace Exploration Agency, \\
3-1-1 Yoshinodai, chuo-ku, Sagamihara, Kanagawa 252-5210}
\altaffiltext{3}{
Department of Physics, Tokyo Metropolitan University,
 1-1 Minami-Osawa, Hachioji, Tokyo 192-0397}
\Received{2011~July~30}
\Accepted{2014~June~4}

\begin{document}
\maketitle

\begin{abstract}
We report properties of the intracluster medium (ICM) in
Abell~1246 to the virial radius ($r_{200}$) and further 
outside as observed with Suzaku.  The ICM emission is 
clearly detected to $r_{200}$, and we derive profiles of 
electron temperature, density, entropy, and cluster mass 
based on the spectral analysis.  The temperature shows 
variation from $\sim 7$ keV at the central region to 
$\sim 2.5$ keV around $r_{200}$.  The total mass in $r_{500}$ 
is $(4.3 \pm 0.4) \times 10^{14}~M_{\odot}$, 
assuming hydrostatic equilibrium. 
At $r>r_{500}$, the hydrostatic mass starts to decline and 
we, therefore, employ the total mass within $r_{200}$ based 
on weak-lens mass profile obtained from a sample of lower 
mass clusters.  This yields the gas mass fraction at $r_{200}$ 
consistent with the cosmic baryon fraction, i.e. $\sim 17$\%. 
The entropy profile indicates a flatter slope than that of 
the numerical simulation, particularly in $r>r_{500}$.  
These tendencies are similar to those of other clusters 
observed with Suzaku.  We detect no significant ICM emission 
outside of $r_{200}$, and $2\sigma$ upper limits of redshifted
O\emissiontype{VII} and O\emissiontype{VIII} line intensities 
are constrained to be less than 2.9 and 
$5.6\times 10^{-7}$ photons cm$^{-2}$ s$^{-1}$ arcmin$^{-2}$, 
respectively.  The O\emissiontype{VII} line
upper limit indicates $n_{\rm H}< 4.7\times 10^{-5}$ cm$^{-3}$
($Z/0.2~Z_{\odot}$)$^{-1/2}$ ($L/20~{\rm Mpc}$)$^{-1/2}$, which
corresponds to an overdensity, $\delta<160$
($Z/0.2~Z_{\odot}$)$^{-1/2}$ ($L/20~{\rm Mpc}$)$^{-1/2}$\@.
\end{abstract}

\section{Introduction}
\label{sec:intro}

Clusters of galaxies, the largest virialized systems in the 
universe, are filled with the intracluster medium (ICM),
which consists of X-ray emitting hot plasma with a typical
temperature of a few times $10^7$ K.  X-ray spectroscopy of
the ICM enables us to determine its temperature and density.  
Clusters are often characterized by the virial radius. 
Within this radius the cluster mass can be determined
under the assumption of hydrostatic equilibrium (H.E.) of the 
ICM, and is a useful parameter for constraining cosmology.
In the framework of a hierarchical structure formation based 
on the cold dark matter paradigm, clusters are thought to grow 
into larger systems through mass accretion flows which are merged
into the ICM at a radius away from a few times the virial 
radius, along large-scale filamentary structures.  The cluster 
outskirts around the virial radius would leave a trace of 
freshly shock-heated accreting matters into the hot ICM. 
In this sense, the outermost edge is the real front of the 
cluster evolution.
However, because of the difficulties in observation, properties, 
such as temperature, density around the virial radius have not 
been known well yet. 

Recent observational studies of clusters with Chandra and 
XMM-Newton, with their powerful imaging capability and large 
effective area, have unveiled radial profiles of temperature, 
entropy, gas mass, and gravitational mass up to $r_{500}$
within which the mean cluster mass density is 500 times the cosmic 
critical density and which is about a half of the virial radius 
\citep{vikhlinin05,piffaretti05,pratt07,vikhlinin09,pratt10,zhang10}.
The derived temperature and entropy profiles to $r_{500}$ are 
almost consistent with theoretical expectations from the 
self-similar assumption. On the other hand, the gas mass fraction 
$M_{\rm gas}/M_{\rm total~mass}$ increases with radius to $r_{500}$, 
and does not exceed the cosmic baryon fraction.

\begin{figure*}[!t]
\begin{center}
\begin{minipage}{0.48\textwidth}
\FigureFile(\textwidth,\textwidth){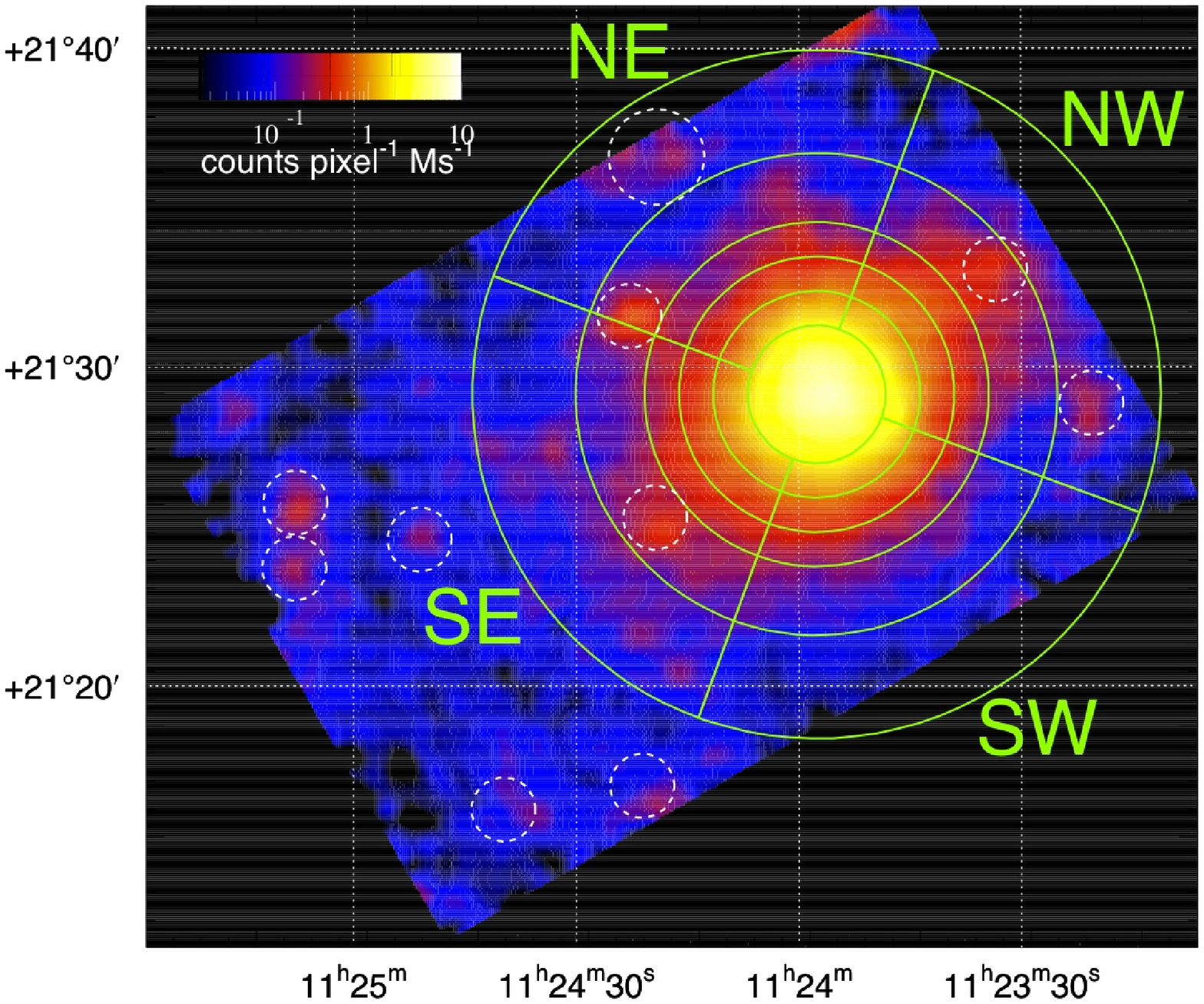}
\end{minipage}
\begin{minipage}{0.48\textwidth}
\FigureFile(\textwidth,\textwidth){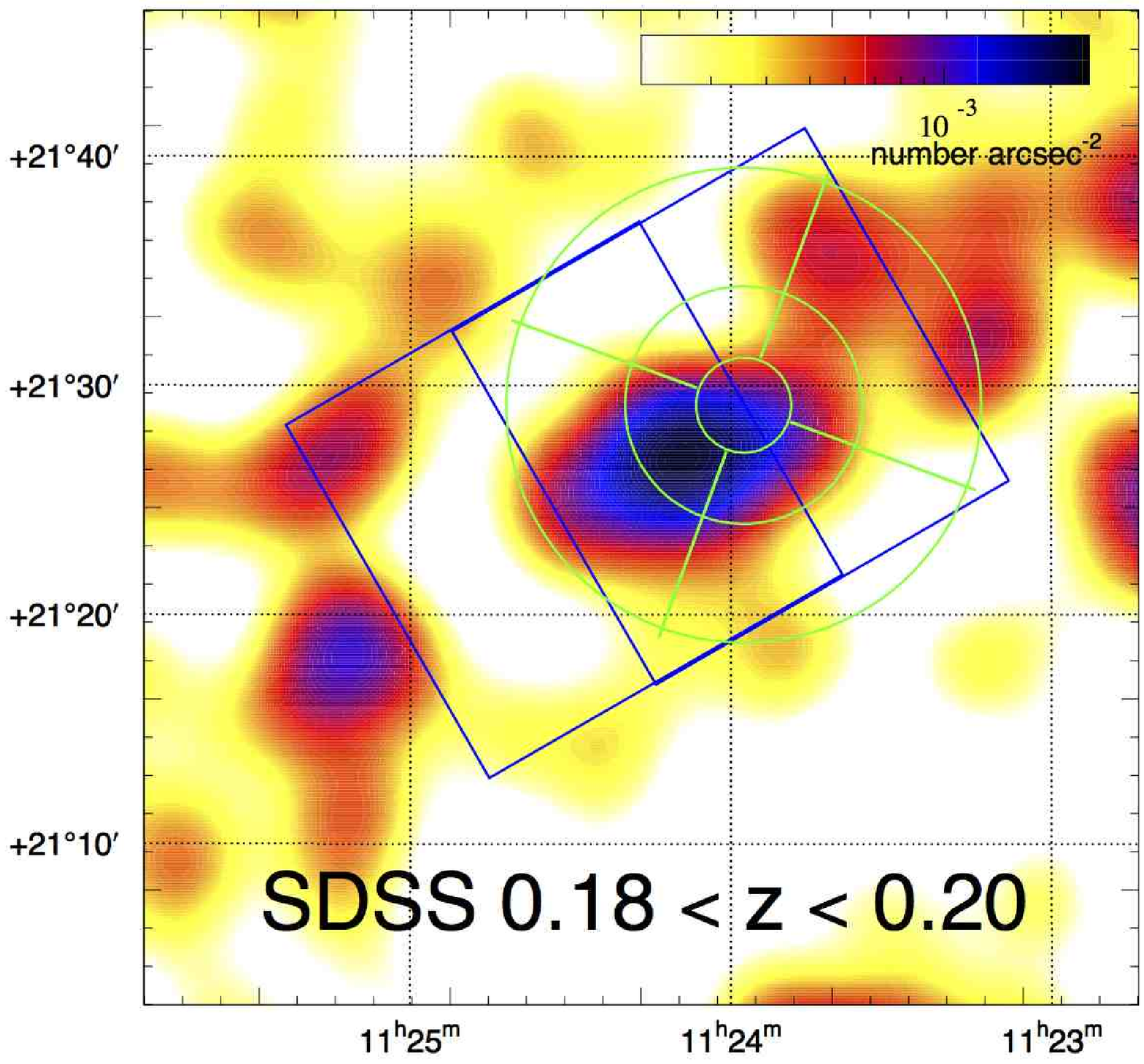}
\end{minipage}

\caption{Left: Combined XIS image of Abell~1246 in 0.5--5.0 keV energy
 range. The observed XIS0, 1, 3 images were added on the sky coordinate
 and smoothed with $\sigma=16$ pixel $\simeq 17''$ Gaussian. The
 instrumental background (NXB) was subtracted and the exposure was
 corrected, although vignetting was not corrected. Green circles show
 the extracted regions for the spectral analysis. The subtracted point
 source regions are indicated by white dashed circles. 
Right: Galaxy number density map from the SDSS catalogue
 around Abell~1246.  The projected galaxy number density was selected 
 between $0.18 < z < 0.20$.  The map consisting of 
 620 galaxies was smoothed with $\sim 2'$
 Gaussian. Blue boxes show the overlapped two Suzaku
 observations. Green circle regions correspond to the
 extracted spectra regions for investigating the directional dependence
 of the ICM properties. For details, please see in text.
}
\label{fig:1}
\end{center}
\end{figure*}

Because Suzaku XIS is characterized by a lower background level
especially above 3~keV and a higher sensitivity below 1~keV 
\citep{koyama07}, we have been able to observe the ICM emission 
beyond $r_{500}$ region of clusters \citep{bautz09, george09, 
reiprich09, hoshino10, kawaharada10, sato10, akamatsu11, ichikawa13}.  
The clusters observed with Suzaku show a similar trend; the 
temperature drops to $\sim$1/3 of the peak from the center to 
the outskirts as expected from structure formation scenarios 
\citep{burns10}. On the other hand, the radial entropy profile 
has a flatter slope compared to that of numerical simulations 
obtained assuming adiabatic cool gas accretion (e.g.\ \cite{voit05}).  
In addition, \citet{kawaharada10} report a directional dependence 
of the temperature and entropy in Abell 1689 cluster which is 
considered to reflect the mass flow from the large scale structure 
filament. Thus, observing the cluster outskirts as a whole, 
not only the limited direction, is important.

In further outer region, as far as the virial radius, the 
intergalactic matter is considered to be not yet mixed with the ICM. 
It is the highest density component of the warm-hot inter galactic 
medium (WHIM, e.g.,\ \cite{cen99,cen06}), which is thought to exist
along the large-scale structure as filaments.  The WHIM would be 
also the most promising candidate for the ``missing baryons'', 
which play key roles for investigating the inconsistency between 
the baryon density in the local universe and the distant universe 
(e.g., \ \cite{fukugita98, rauch98, takei11}).
Although it is difficult to detect the WHIM with current detectors 
such as CCD cameras (e.g.\ Suzaku XIS), if observed, its thermal 
and chemical properties would provide rich information 
on the structure formation and evolution of the universe. 
Some challenging observations have given a constraint on the upper 
limit of the WHIM emission. High resolution imaging spectroscopy 
with Chandra and XMM-Newton is claimed to show evidences for the 
WHIM emission (e.g., \cite{kaastra03,werner08,galeazzi09}).
Their grating observations also make the absorption-line 
study and restrict the WHIM density significantly
(e.g., \cite{nicastro05,kaastra06,buote09,fang10,zappacosta10}).
Recent studies with Suzaku have shown that Suzaku XIS can effectively 
constrain the Galactic emission \citep{gupta09,yoshino09} better 
than previous satellites. Therefore, Suzaku also has a great advantage 
in the WHIM search, because a reliable estimation of the foreground 
Galactic emission is of utmost important in constraining the WHIM emission.
Although \citet{takei07} detect no significant redshifted O lines 
from the outer region of Abell~2218 with Suzaku, they set a strict 
constraint on the intensity 
(see also \cite{tamura08,sato10,mitsuishi12}).

\begin{table*}[t]
\caption{Suzaku observation logs of Abell~1246 cluster.}
\label{tab:1}
\begin{tabular}{lccccc} \hline 
Region & Seq. No. & Obs. date & \multicolumn{1}{c}{(RA, Dec)$^\ast$} &Exp.&After screening \\
&&&J2000& ksec &(BI/FI) ksec \\
\hline 
Abell 1246 & 804028010 & 2009-11-16T04:54:20 & (\timeform{11h23m59.8s},
 \timeform{+21D29'11''})& 48.5& 36.9/36.9\\
Abell 1246 offset & 804029010 & 2009-11-28T00:30:33 & (\timeform{11h24m30.7s},
 \timeform{+21D25'09''})& 80.4& 52.1/52.1\\
\hline\\[-1ex]
\multicolumn{6}{l}{\parbox{0.9\textwidth}{\footnotesize 
\footnotemark[$\ast$]
Average pointing direction of the XIS, written in the 
RA\_NOM and DEC\_NOM keywords of the event FITS files.}}\\
\end{tabular}
\end{table*}

Abell~1246 is a cluster of galaxies characterized by a smooth 
distribution of the ICM.  ASCA observation determined the 
temperature, $kT$ and the metal abundance, 
$Z$, as $5.17\pm0.58$ keV and $0.26\pm0.17$ solar, respectively, 
averaged over the whole cluster \citep{fukazawa04}. 
The redshift of Abell~1246 cluster is 0.1902 from 
the NASA/IPAC Extragalactic Database\footnote{ 
http://ned.ipac.caltech.edu/}. 
At this redshift, $1'$ corresponds to 191~kpc.
Here, we use $H_0=70$ km~s$^{-1}$~Mpc$^{-1}$, 
$\Omega_{\Lambda} = 1-\Omega_M = 0.73$.
The virial radius, $r_{200}$, 
using the mean temperature, was defined as 
$r_{200} = 2.77 ( \langle kT \rangle /10~{\rm keV})^{1/2}/E(z)$ Mpc 
and  $E(z)=(\Omega_M (1+z)^3+ \Omega_{\Lambda})^{1/2}$
as described in \citet{henry09}, in this paper. 
For Abell 1246 cluster, the virial radius, $r_{200}$, is 
1.97 Mpc or \timeform{10.3'} with $\langle kT \rangle = 6$ keV.
Throughout this paper we adopt the Galactic hydrogen column 
density of $N_{\rm H} = 1.57\times 10^{20}$ cm$^{-2}$ \citep{dickey90} 
in the direction of Abell~1246, and use the solar abundance 
table provided by \citet{anders89}.  Unless noted otherwise, 
the errors are in the 90\% confidence region for a single 
interesting parameter.

\section{Observations}
\label{sec:obs}

Suzaku carried out Abell~1246 and its offset observation in 
November 2009 (PI: K. Sato) to observe the outskirts of the 
cluster beyond the virial radius or $r_{200}$.
The observation logs are shown in table~\ref{tab:1}, 
and the XIS image in the 0.5--5.0~keV energy range is shown in 
figure~\ref{fig:1}.  The XIS was operated in the normal clocking 
mode (8~s exposure per frame), in the standard $5\times 5$ or 
$3\times 3$ editing mode.  During these observations, a significant 
effect of the Solar Wind Charge exchange (SWCX) was not confirmed 
in ACE data \footnote{http://www.srl.caltech.edu/ACE/ASC/}.
We note that Abell~1246 observations have an attitude uncertainty, 
which are estimated within $\sim1$ arcmin, because of a satellite 
house keeping system problem.  We, however, conclude that the 
uncertainty is smaller by visual inspection, and it does not 
affect our results.

\begin{figure}[thp]
\centerline{\FigureFile(0.45\textwidth,8cm){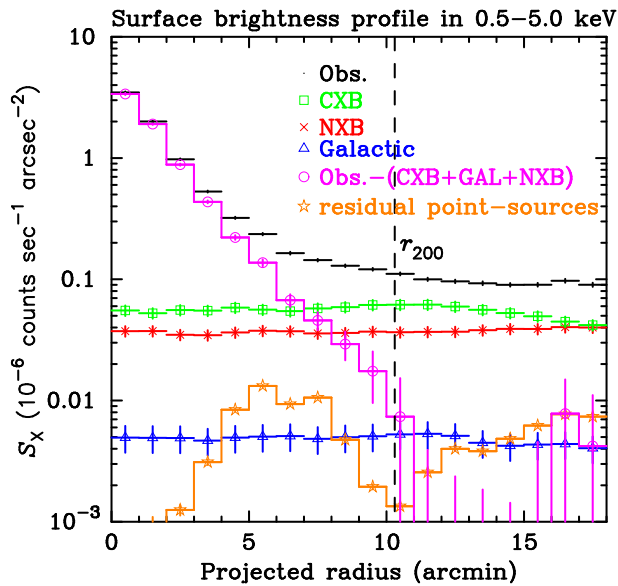}}
\vspace*{3ex}
\caption{Radial surface brightness profile in 0.5--5.0 keV\@. Point-like sources
specified in figure \ref{fig:1} (left) are removed, but vignetting is
not corrected. Observed data profile (black) is shown with the CXB
 (green), NXB (red), and Galactic components (blue) profiles. 
A resultant background-subtracted profile (black$-$green$-$red$-$blue) 
is shown in magenta. The error bars in these profiles are 1$\sigma$\@.
In the error bars of the background-subtracted profile (magenta), 
the uncertainties of the backgrounds are added in quadrature to the 
corresponding statistical 1$\sigma$ errors.
}
\label{fig:2}
\end{figure}

\section{Data Reduction}
\label{sec:data}

We used version 2.4 processing Suzaku data, and the analysis 
was performed with HEAsoft version 6.10 and XSPEC 12.6.0q.
In the analysis of the XIS data, we selected ELEVATION $>$ 15$^\circ$ 
of the standard data set
\footnote{http://www.astro.isas.ac.jp/suzaku/process
/v2changes/criteria\_xis.html}
to remove stray-light from the day Earth limb.
Event screening with cut-off rigidity (COR) of ``COR2 $>$ 8'' 
was also performed in our data.  The exposure after the screening 
is shown in table \ref{tab:1}. 

In order to subtract the non-X-ray background (NXB), 
we employed the dark Earth database using the ``xisnxbgen'' Ftools task.
For spectral fits of the ICM emission, we generated ancillary 
response files (ARFs) for Abell~1246, assuming the $\beta$-model 
surface brightness profile as $\beta =0.52 $ and $r_c =$
\timeform{0.47'} \citep{fukazawa04} by ``xissimarfgen'' \citep{ishisaki07}.
We also generated ARFs for each observation and  
assumed a uniform sky of \timeform{20'} radius for the Galactic and 
Cosmic X-ray Background (CXB) emissions.  We included the effect 
of the contaminations on the optical blocking filter of the XISs 
in the ARFs.  Because the energy resolution slowly degraded
after the launch due to radiation damage, this effect was included 
in redistribution matrix file (RMF) by ``xisrmfgen'' Ftools task.

\section{Spectral Analysis}

\begin{figure*}[!th]
\begin{minipage}{0.45\textwidth}
\FigureFile(\textwidth,\textwidth){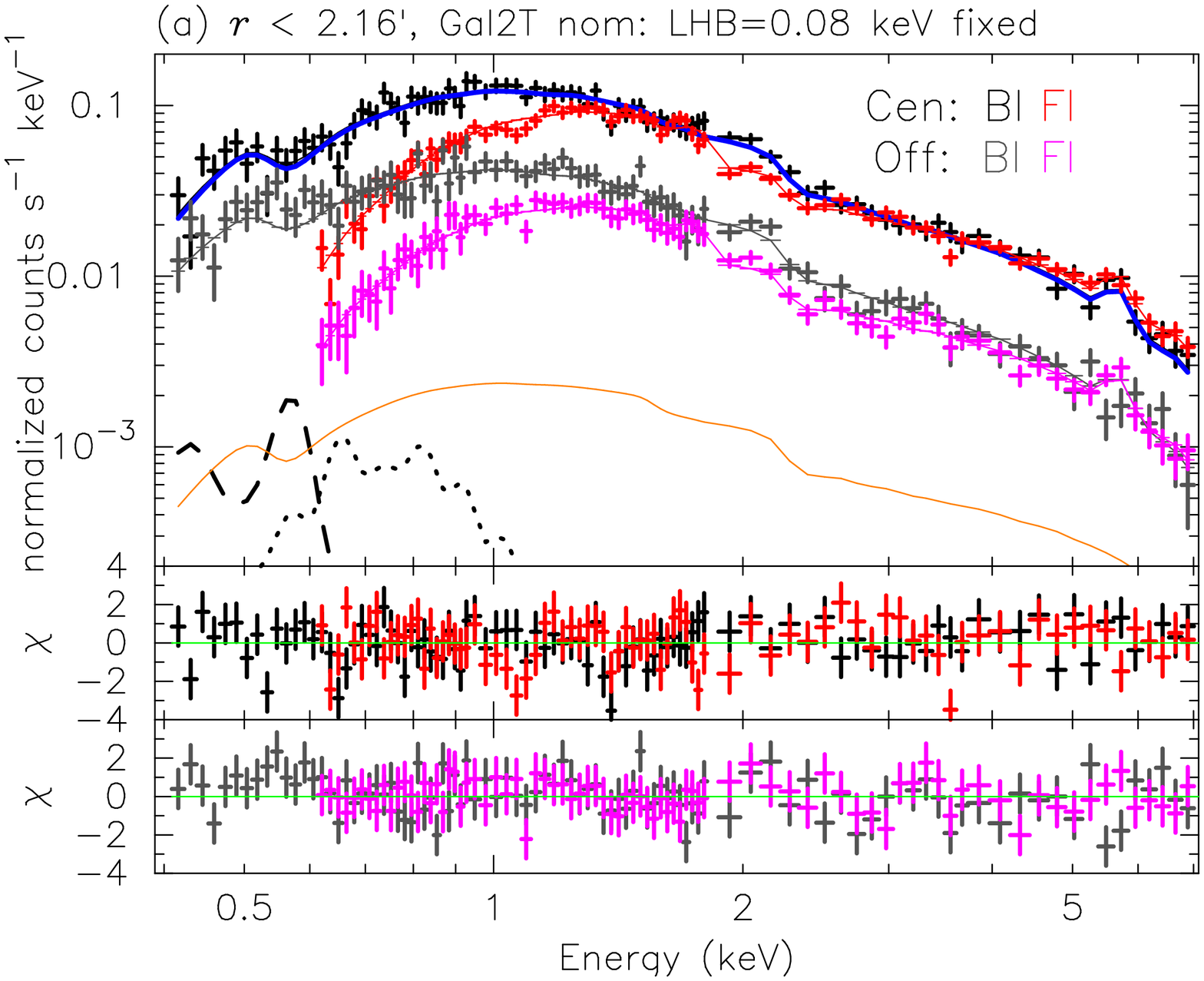}
\end{minipage}\hfill
\begin{minipage}{0.45\textwidth}
\FigureFile(\textwidth,\textwidth){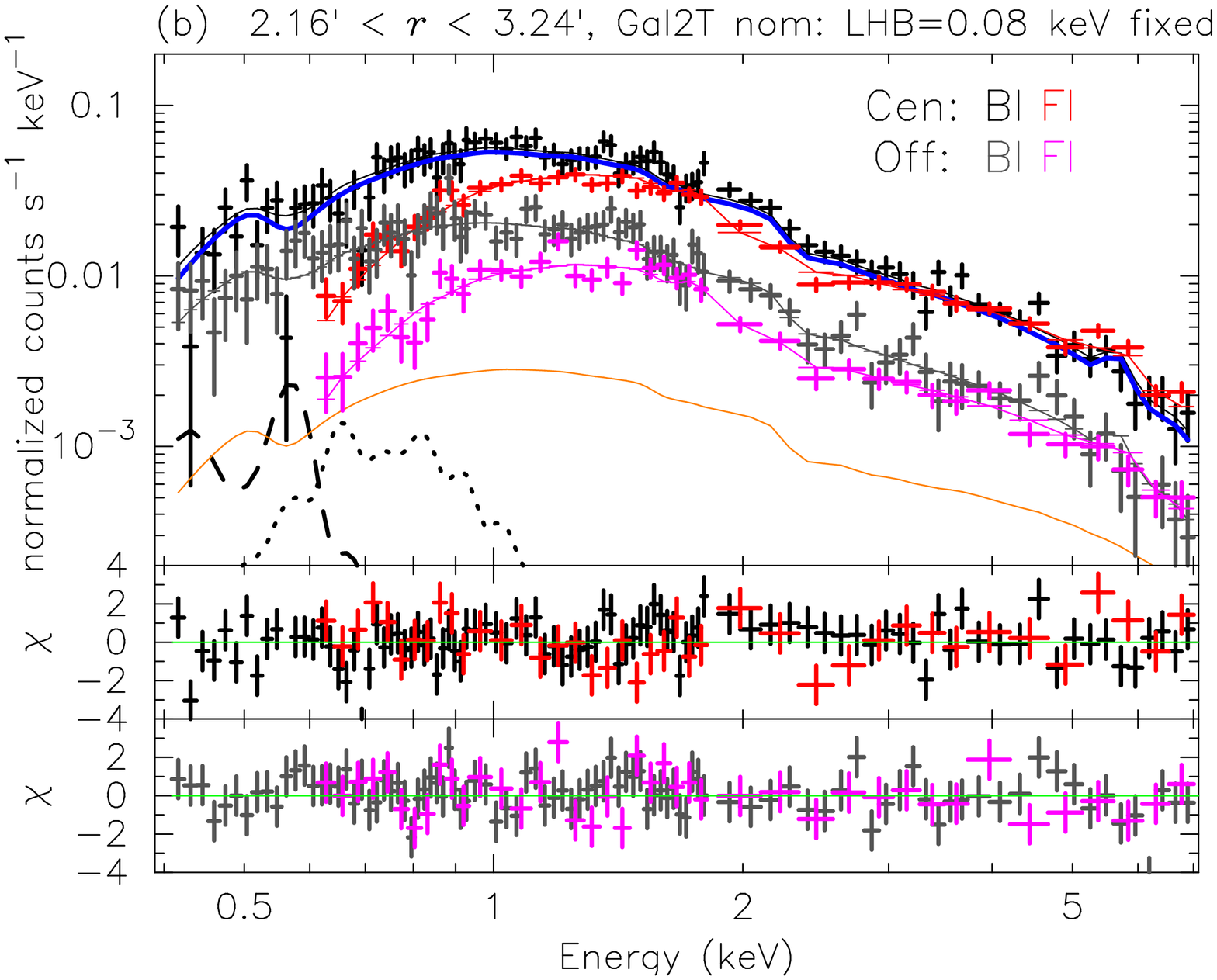}
\end{minipage}

\begin{minipage}{0.45\textwidth}
\FigureFile(\textwidth,\textwidth){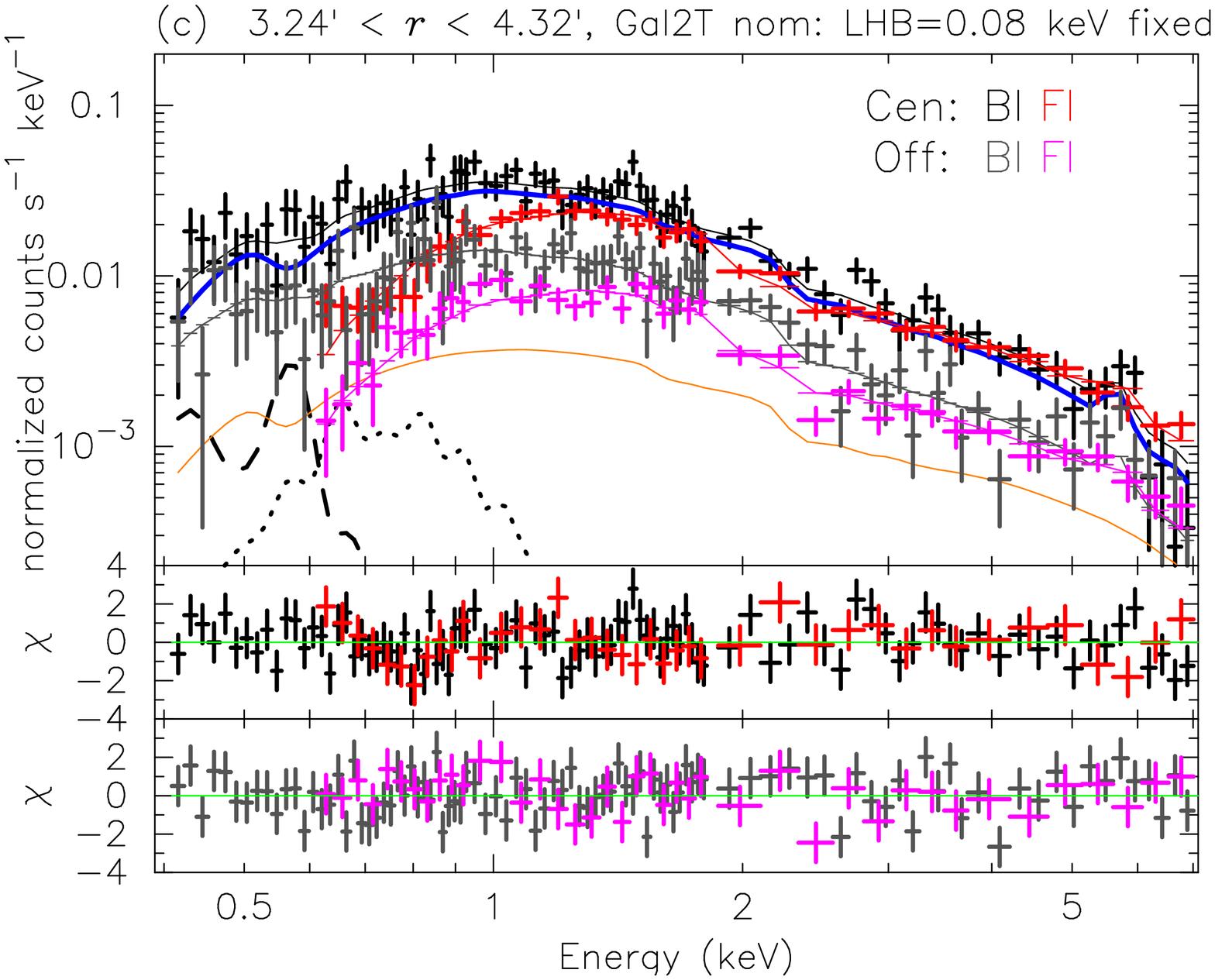}
\end{minipage}\hfill
\begin{minipage}{0.45\textwidth}
\FigureFile(\textwidth,\textwidth){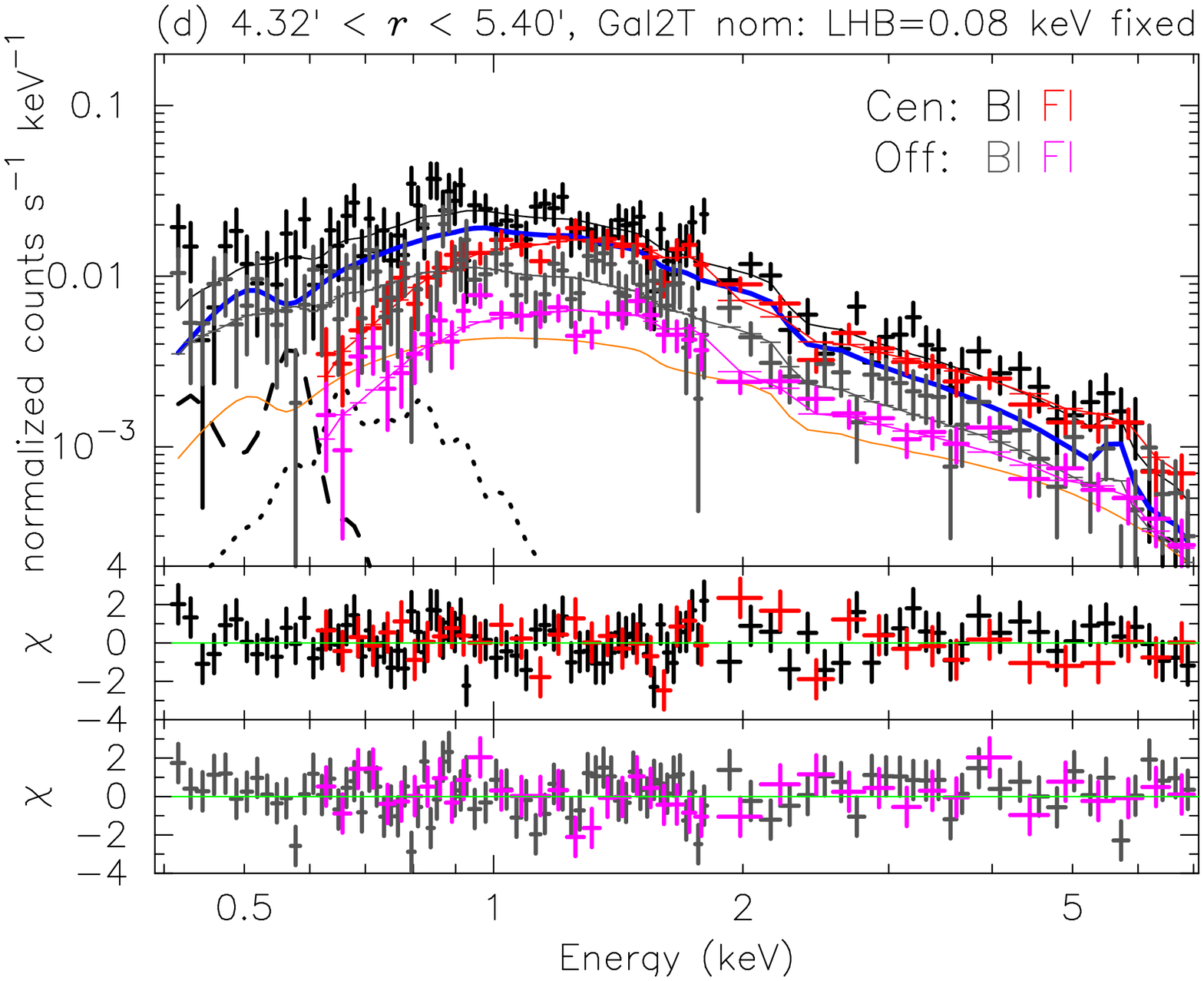}
\end{minipage}
\vspace*{3ex}
\caption{
The panels form (a) to (g) show the observed spectra sorted in
the seven annular regions. Spectra obtained from the BI and FI CCDs of the
central pointing data are presented in black and red, respectively,
after subtracting only NXB. Those of the offset pointing data are in
dark gray and magenta, as well. Blue lines indicate the ICM component 
for each annular region of (a)--(f).
The LHB, MWH and CXB components for the BI spectra are shown 
in black-dashed, black-dotted, and orange-solid lines, respectively.
The energy range around the Si K-edge (1.825--1.840 keV) is ignored
in the spectral fits.
The lower panels show the fit residuals in units of $\sigma$.
}
\label{fig:3}
\end{figure*}
\addtocounter{figure}{-1}
\begin{figure*}[!th]
\begin{minipage}{0.45\textwidth}
\FigureFile(\textwidth,\textwidth){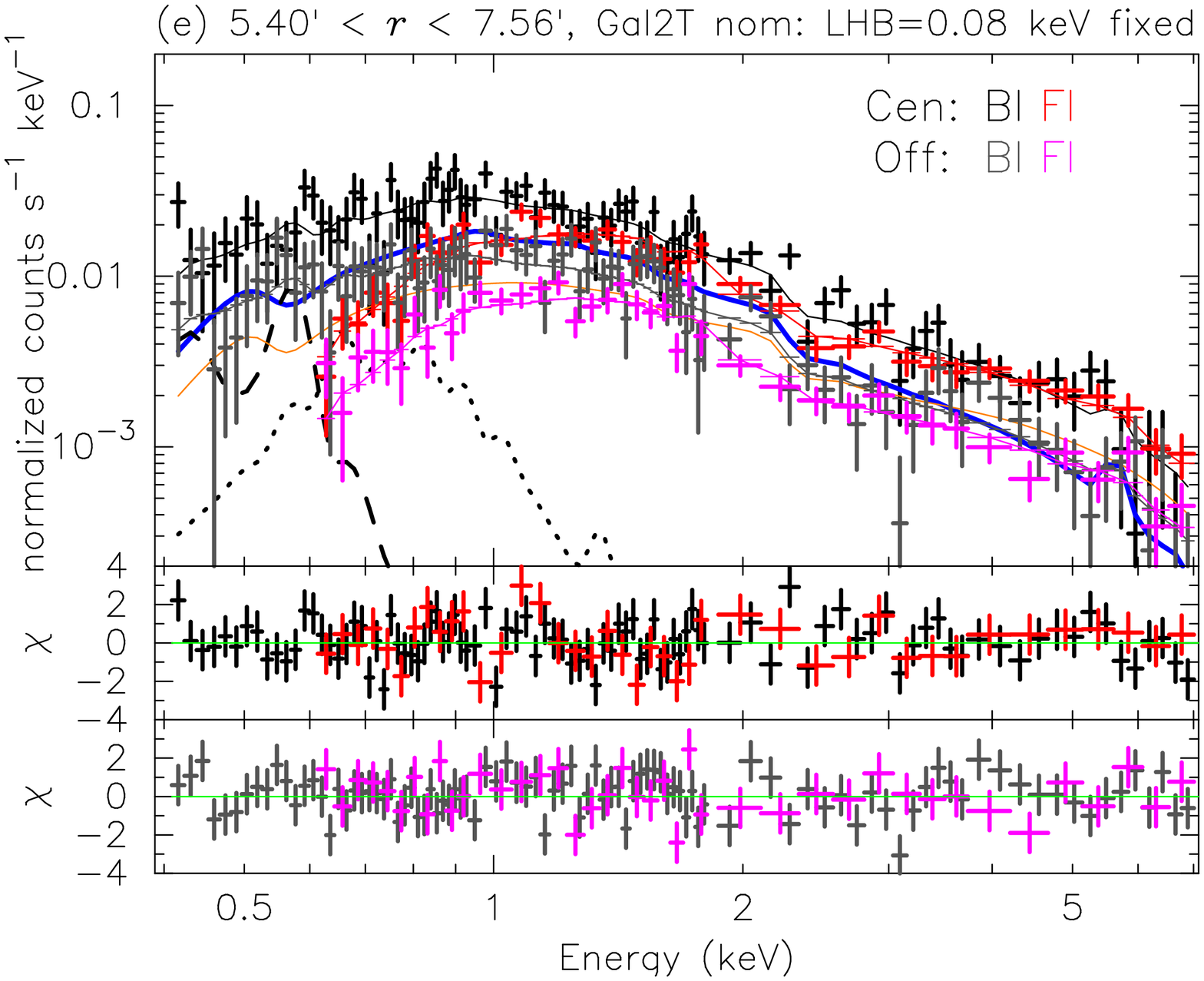}
\end{minipage}\hfill
\begin{minipage}{0.45\textwidth}
\FigureFile(\textwidth,\textwidth){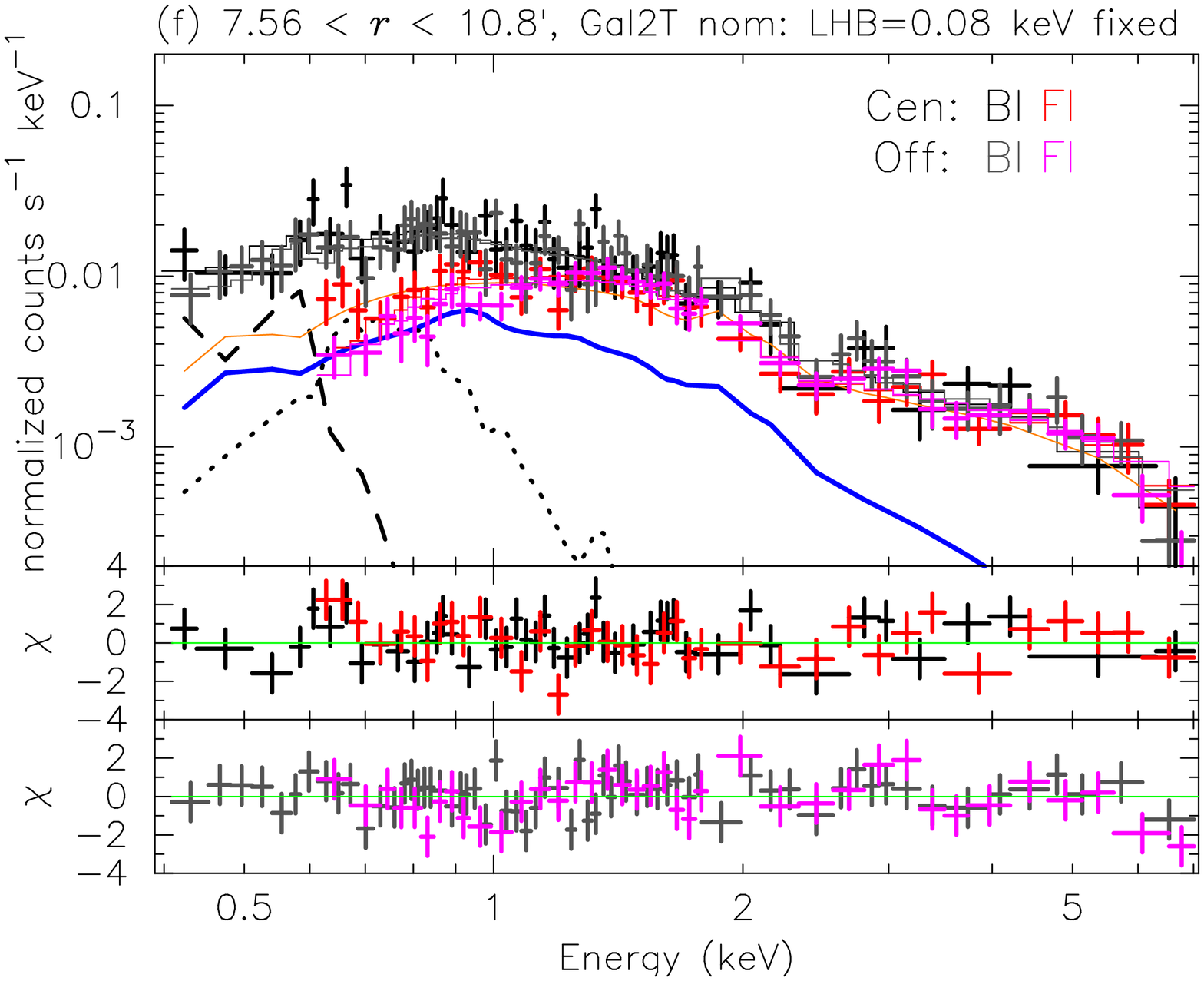}
\end{minipage}

\begin{minipage}{0.45\textwidth}
\FigureFile(\textwidth,\textwidth){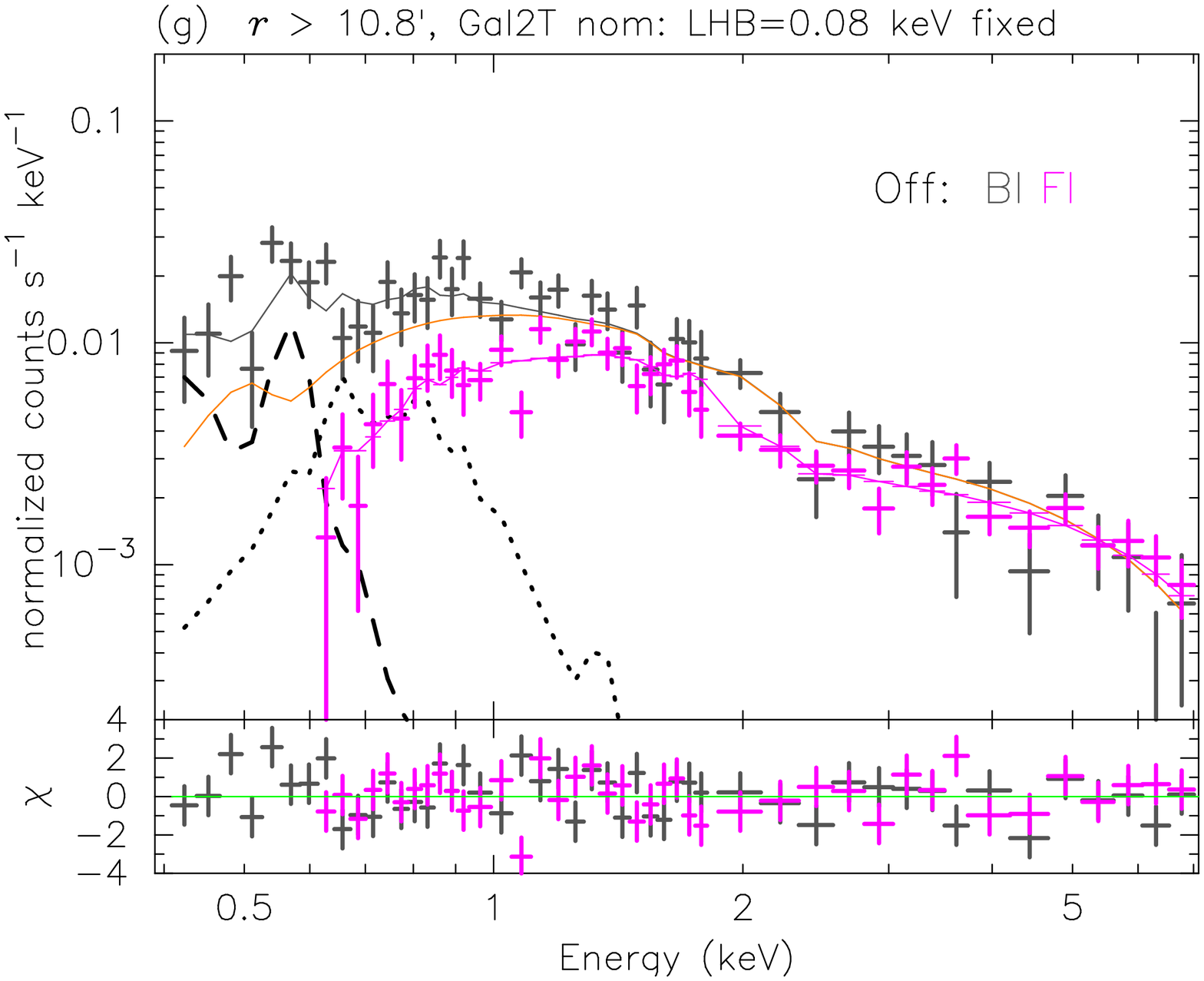}
\end{minipage}\hfill
\begin{minipage}{0.45\textwidth}
\end{minipage}
\vspace{3ex}
\caption{
continue.
}
\label{fig:3-2}
\end{figure*}

\subsection{Subtraction of point-like sources}
\label{subsec:pointsource}

Using ``wavdetect'' tool in CIAO
\footnote{http://cxc.harvard.edu/ciao/}, 
we searched point-like sources in the Suzaku images in the energy 
range of 0.5--2.0 and 2--10 keV which were useful energy bands 
for detecting galaxies or galaxy groups and point-like sources, 
respectively. The ``wavdetect'' tool in CIAO is based on scalable, 
oscillatory functions which deviate from zero only within 
a limited spatial regime and have average value zero 
\citep{freeman02}. This tool, therefore, is useful for 
characterizing simultaneously the shape, location, and strength 
of astronomical sources. In addition, this algorithm operates 
effectively regardless of the PSF shape. Because the tool does 
not deconvolve the Suzaku's PSF, we regard the sources which are 
detected by the algorithm and smaller than the Suzaku's PSF as 
the compact diffuse or point-like sources. As shown by the white 
dashed circles in figure~\ref{fig:1}, we detected 10 sources as 
compact diffuse sources or point-like sources with the significance 
threshold of 3 $\sigma$. The same 10 sources were also identified 
when we used another tool, ``wavelet'' in the SAS 
package\footnote{http://xmm.esa.int/sas/}. 
We subtracted 9 point-like sources with 1 arcmin and another source 
with 1.5 arcmin. We checked fluxes of all the detected sources with 
a power-law model of the photon index to be fixed at 1.4\@. The flux 
of the faintest point-like source in 2--10 keV is 
$\sim 3~\times 10^{-14}$ erg cm$^{-2}$ s$^{-1}$, and we estimate the 
CXB level and fluctuations with the value. Figure \ref{fig:2} shows 
0.5--5.0 keV radial surface brightness profile (background inclusive, 
but point-like or compact sources excluded) as black crosses. 
As shown in \citet{kawaharada10}, because half of the point-like 
source signals is expected to escape from the masked  1$'$ radius 
region due to the PSF, we need to estimate the residual signals for 
the systematic uncertainties of the background as the contaminated 
signal. We simulated the residual signals from point sources, the 
Galactic and CXB emissions using ``xissim'' tool \citep{ishisaki07} 
with 10 times longer exposure time than actual observations. 
Here, the Galactic and CXB level were assumed to be the flux level 
of the default case in table \ref{tab:2}. The estimated 
contaminations from the point-like or compact sources are shown in 
figure \ref{fig:2} in the orange line. In our analysis, the 
contributions from the residual signals and unresolved sources are 
included in the CXB model. In $r>$\timeform{10.8'}, the uncertainties 
of the flux are $\sim 8$\% comparing to the estimated CXB level.

\begin{table*}[htb]
\caption{Summary of the parameters of the fits for model I in
the $r>$\timeform{10.8'} region owing to background estimation.}
\label{tab:2}
\begin{center}
\begin{tabular}{lcccc} \hline 
\hline
\multicolumn{5}{c}{model I} \\
\hline
%Region  & $r>$\timeform{10.8'}  & case~1 & case~2 & case~3 \\
 & & LHB/CXB free & LHB free & default \\
%\hline
 & Const. (BI/FI) & 1(fixed)/0.97$^{+0.05}_{-0.04}$ & 1(fixed)/0.98$^{+0.08}_{-0.08}$ & 1(fixed)/0.97$^{+0.08}_{-0.08}$ \\ 
\hline
\multicolumn{2}{l}{Galactic \& CXB} & & & \\
\hline
LHB & $kT$ (keV) & 0.07$^{+0.05}_{-0.07}$ & 0.07$^{+0.05}_{-0.07}$ & 0.08(fixed)\\
 & $Norm^{\ast}$ ($\times10^{-3}$) & 10.39$^{+0.25}_{-0.29}$ &
	     12.12$^{+0.28}_{-0.28}$ & 6.05$^{+1.41}_{-1.57}$ \\
MWH & $kT$ (keV) & 0.28$^{+0.05}_{-0.05}$ & 0.29$^{+0.05}_{-0.05}$ & 0.29$^{+0.06}_{-0.05}$ \\
 & $Norm^{\ast}$ ($\times10^{-4}$) & 2.45$^{+0.74}_{-0.77}$ &
	     2.63$^{+0.76}_{-0.73}$ & 2.51$^{+0.81}_{-1.07}$ \\
CXB  & $\Gamma$ & 1.45$^{+0.07}_{-0.07}$ & 1.4(fixed) & 1.4(fixed)\\ 
% & $Norm^{\ast}$ & 9.90$^{+0.89}_{-0.86}$ &
%	     9.50$^{+0.68}_{-0.63}$ & 9.50$^{+0.67}_{-0.62}$ \\
 & $S_{X}^{\dagger}$ & 6.13$^{+0.55}_{-0.53}$ &
	     5.88$^{+0.42}_{-0.39}$ & 5.88$^{+0.41}_{-0.38}$ \\
\hline
\multicolumn{2}{c}{$\chi^2$/d.o.f.$^{\dagger\dagger}$} & 1965/1771 & 1966/1772 & 1966/1773\\
\hline\\[-1ex]
\multicolumn{5}{l}{\parbox{0.8\textwidth}{\footnotesize 
\footnotemark[$\ast$] 
Normalization of the {\it apec} component
divided by the solid angle, $\Omega^{\makebox{\tiny\sc u}}$,
assumed in the uniform-sky ARF calculation (20$'$ radius),
${\it Norm} = \int n_{\rm e} n_{\rm H} dV \,/\,
(4\pi\, (1+z)^2 D_{\rm A}^{\,2}) \,/\, \Omega^{\makebox{\tiny\sc u}}$
$\times 10^{-14}$ cm$^{-5}$~400$\pi$~arcmin$^{-2}$, 
where $D_{\rm A}$ is the angular distance to the source.}}\\
\multicolumn{5}{l}{\parbox{0.8\textwidth}{\footnotesize 
\footnotemark[$\dagger$] 
The 2--10 keV CXB surface brightness in units of 
$\times 10^{-8}$ erg cm$^{-2}$ s$^{-1}$ sr$^{-1}$.
}}\\
\multicolumn{5}{l}{\parbox{0.8\textwidth}{\footnotesize 
\footnotemark[$\dagger\dagger$] 
The $\chi^2$ values show total values of simultaneous fits of the 
ICM and background components.
}}\\
\end{tabular}
\end{center}
\end{table*}

\begin{table*}[htb]
\caption{Summary of the ICM parameters of the fits for model I: default.}
\label{tab:3}
\begin{center}
\begin{tabular}{lcccc} \hline 
\hline
\multicolumn{5}{c}{model I: default} \\
\hline
Region  &  & $r<$\timeform{2.16'} & \timeform{2.16'} $<r<$
 \timeform{3.24'} & \timeform{3.24'} $<r<$ \timeform{4.32'} \\
\hline
Center & Const. (BI/FI) & 1(fixed)/0.98$^{+0.02}_{-0.02}$ &
	     1(fixed)/1.02$^{+0.04}_{-0.04}$ & 1(fixed)/1.02$^{+0.05}_{-0.05}$ \\
Offset & Const. (BI/FI) & 1.13$^{+0.04}_{-0.04}$/0.94$^{+0.03}_{-0.03}$
	 & 1.07$^{+0.05}_{-0.05}$/0.94$^{+0.05}_{-0.04}$ & 1.07$^{+0.06}_{-0.06}$/0.94$^{+0.06}_{-0.06}$ \\ 
\hline
 & $kT$ (keV) & $^\dagger$6.80$^{+0.41~+0.01}_{-0.41~-0.01}$ & 6.36$^{+0.42~+0.04}_{-0.41~-0.05}$ &  6.17$^{+0.60~+0.10}_{-0.57~-0.11}$ \\
 & $Z$ (solar) & 0.20$^{+0.05~+0.00}_{-0.05~-0.00}$ & 0.18$^{+0.09~+0.00}_{-0.09~-0.00}$ &
		 0.24$^{+0.15~+0.00}_{-0.14~-0.01}$ \\
 & $Norm^{\ast}$ & 2.59$^{+0.06+0.01}_{-0.06~-0.03} \times 10^{-4} $ &
	     3.40$^{+0.12~+0.02}_{-0.12~-0.02} \times 10^{-5}$ &
		 2.24$^{+0.12~+0.03}_{-0.12~-0.03} \times 10^{-5} $ \\
\hline
\hline
Region & & \timeform{4.32'} $<r<$ \timeform{5.40'} & \timeform{5.40'}
	     $<r<$ \timeform{7.56'} & \timeform{7.56'} $<r<$ \timeform{10.8'} \\
\hline
Center & Const. (BI/FI) & 1(fixed)/1.02$^{+0.07}_{-0.06}$ &
	     1(fixed)/1.02$^{+0.08}_{-0.07}$ &
		 1(fixed)/1.04$^{+0.11}_{-0.10}$ \\
Offset & Const. (BI/FI) & 1.07$^{+0.08}_{-0.08}$/0.94$^{+0.07}_{-0.07}$
	 & 1.07$^{+0.08}_{-0.08}$/0.94$^{+0.07}_{-0.06}$ &
		 1.06$^{+0.09}_{-0.08}$/0.91$^{+0.08}_{-0.07}$  \\ 
\hline
 & $kT$ (keV) & 5.06$^{+0.71~+0.16}_{-0.56~-0.18}$ & 4.20$^{+0.71~+0.32}_{-0.60~-0.33}$ &
		 2.37$^{+1.16~+0.76}_{-0.90~-0.71}$  \\
 & $Z$ (solar) & 0.24$^{+0.21~+0.00}_{-0.19~-0.01}$ & 0.24$^{+0.27~+0.03}_{-0.24~-0.03}$ &
		 0.13$^{+0.43~+0.00}_{-0.13~-0.03}$  \\
 & $Norm^{\ast}$ & 1.19$^{+0.09~+0.02}_{-0.09~-0.03} \times 10^{-5}$ &
	     5.38$^{+0.62~+0.22}_{-0.62~-0.17} \times 10^{-6} $ &
		 1.60$^{+0.55~+0.17}_{-0.49~-0.05} \times 10^{-6}  $ \\
\hline
\hline\\[-1ex]
\multicolumn{5}{l}{\parbox{0.8\textwidth}{\footnotesize
\footnotemark[$\ast$] 
Normalization of the {\it vapec} component scaled with a factor of
the selected region comparing to the assumed image in
 ``xissimarfgen'',
${\it Norm}= {factor} \int
n_{\rm e} n_{\rm H} dV \,/\, [4\pi\, (1+z)^2 D_{\rm A}^{\,2}]$ $\times
10^{-14}$~cm$^{-5}$~arcmin$^{-2}$, where $D_{\rm A}$ is the angular
distance to the source. }}\\
\multicolumn{5}{l}{\parbox{0.8\textwidth}{\footnotesize
\footnotemark[$\dagger$] 
The first and second errors correspond to 
the statistical error and the systematic errors by changing 
the CXB level by $\pm 10$\%, respectively. }}\\
\end{tabular}
\end{center}
\end{table*}

\subsection{Simultaneous spectral fits for all regions}
\label{subsec:eachspectra}

We extracted spectra from seven annular regions of 
$r<$\timeform{2.16'}, 
\timeform{2.16'}$<r<$\timeform{3.24'},
\timeform{3.24'}$<r<$\timeform{4.32'},
\timeform{4.32'}$<r<$\timeform{5.40'}, 
\timeform{5.40'}$<r<$\timeform{7.56'},
\timeform{7.56'}$<r<$\timeform{10.8'}, 
and $r>$\timeform{10.8'}, 
centered on (\timeform{11h23m57.6s}, \timeform{+21D29'09''}).
The annular spectra for each observation are shown in figure
\ref{fig:3}. The spectra of both BI and FI for all regions as shown 
in figure \ref{fig:3} (a)--(g), were fitted simultaneously 
in the energy range of 0.4--7.1/0.6--7.1 keV (BI/FI) for the central 
and offset observations.  In the simultaneous fit, the common
Galactic emission and CXB components were included for all the regions.  
We excluded the narrow energy band around the Si K-edge
(1.825--1.840 keV) because its response was not modeled correctly.  
The energy range below 0.4 keV was also excluded because the C edge 
(0.284 keV) seen in the BI spectra could not be reproduced well in 
our data.  The range above 7.1 keV was also ignored because the Ni line
($\sim7.5$ keV) in the background left a spurious feature after the 
NXB subtraction at large radii.  In the simultaneous fits of the BI 
and FI data, only the normalization parameter was allowed to vary.

We assumed the CXB and two Galactic emissions, local hot bubble 
(LHB) and milky way halo (MWH) as the back- and fore- ground emissions 
in $r>$\timeform{10.8'} and a thermal (ICM) model for the inner region 
of the virial radius as follows (hereafter model I);
\begin{itemize}
\item model I: 
apec$_{\rm LHB}+{\rm phabs}\times({\rm apec}_{\rm MWH}+{\rm apec}_{r<10.8'}+{\rm pow}_{\rm CXB})$
\end{itemize} 

Although there was no flare in the solar-wind flux during our 
observation, because the SWCX and LHB could be hardly distinguished 
with a limited spectral resolution of CCDs
(e.g., \cite{yoshino09, gupta09}), we modeled the sum of the SWCX and 
the LHB as a single unabsorbed thermal plasma. Here we assumed a zero
redshift and 1 solar metallicity for the LHB and MWH emissions.  

We examined the spectral fits by changing the parameters of the LHB 
temperature and the photon index of CXB to be either free or fixed 
as shown in table \ref{tab:2}.
Consequently, as shown in table \ref{tab:2} and \ref{tab:3}, and 
figure \ref{fig:2} and \ref{fig:3}, the ICM emission is significantly 
detected to the virial radius, and the observed spectra are 
well-represented by the model I in which the LHB temperature and the 
photon index of CXB are fixed to be 0.08~keV and 1.4, respectively.

The derived surface brightness of the CXB component in 2--10 keV is 
$5.88^{+0.41}_{-0.38} \times 10^{-8}$ erg cm$^{-2}$ s$^{-1}$ sr$^{-1}$ 
as shown in table~\ref{tab:2}, and the estimated CXB fluctuation 
for the $r>$\timeform{10.8'} region is 12\%. 
The CXB surface brightness agrees with
that of \citet{ichikawa13}, $5.17^{+0.26}_{-0.23} \times 10^{-8}$
erg cm$^{-2}$ s$^{-1}$ sr$^{-1}$ 
(after subtraction of point sources brighter than 
$2 \times 10^{-14}$ erg cm$^{-2}$ s$^{-1}$), 
within statistical errors taking into account the CXB fluctuation 
and the residual flux from the extracted point-like sources.
Using the same threshold, the CXB surface brightness derived with 
previous Suzaku observations is
4--6 $\times 10^{-8}$ erg cm$^{-2}$ s$^{-1}$ sr$^{-1}$
(e.g.,\ \cite{kawaharada10, hoshino10}), they also agree with 
our CXB estimation. The details of the CXB estimation are shown 
in Appendix \ref{app:1}. Because the ICM component shape around the 
outskirts is far from a power-law model one with a photon index of 
1.4 for the CXB component, the effects from the CXB contamination 
would be negligible for our discussions.

We also examined whether or not an additional apec component
(with a redshift of either zero or a cluster value) significantly 
improved $\chi^2$ against a change in the degree
of freedom $\delta \nu$ in the outer region of the virial radius. 
we assumed an additional thermal model for the outer 
region of the virial radius as the Galactic emission (model II) and
ICM emission (model III) as follows, 
\begin{itemize}
\item model II:
apec$_{\rm LHB}+{\rm phabs}\times({\rm apec}_{\rm MWH}+{\rm apec}_{r<10.8'}+{\rm apec}_{r>10.8',~Z=1,~z=0}+{\rm pow}_{\rm CXB})$

\item model III:
apec$_{\rm LHB}+{\rm phabs}\times({\rm apec}_{\rm MWH}+{\rm apec}_{\rm all~regions}+{\rm pow}_{\rm CXB})$

\end{itemize} 
In case of model II, which corresponds to the 
zero redshift apec model, the improvement over the values 
shown in table \ref{tab:2} is $\chi^2=4$ 
with $\delta \nu =2$, while the resultant temperature 
of the additional model is 0.62$^{+0.18}_{-0.16}$ keV\@.  
This is not significant in F-test.
On the other hand, in the case of model III, an addition of a 
cluster-redshift apec model ($z=0.1902$) results in no improvement 
($\delta \chi^2<1$), while the temperature of the 
additional component is $\sim1$ keV, which is comparable to the 
temperature reported for the WHIM emission (e.g., \cite{werner08}).

\subsection{Systematic Errors}
\label{subsec:systematic}

We investigated the effect of a possible incorrect calibration,
such as NXB level and contaminations on XIS optical blocking filter
(OBF), by artificially changing these values by $\pm$10\% and 
comparing the resultant $\chi^2$ value.  While the temperatures of 
the LHB and MWH did not change within $<1$\% compared to the values 
in table~\ref{tab:2}, the normalization of the CXB model changed 
by $\sim$10\%. When we examined the other uncertainty 
in the OBF contaminant by changing the absorber thickness by $\pm10$\%, 
the temperatures of the LHB and MWH did not change within $<$1\%, 
and the normalizations of the LHB, MWH, and CXB changed by $\sim$5\%. 
We also changed the CXB levels by $\pm 10$\% and $\pm 20$\% for the 
fits of the azimuthal average and directional dependence, 
respectively, corresponding to the estimated CXB fluctuation as 
mentioned in Appendix \ref{app:1}.  Even if the CXB level of the 
outermost region for the azimuthal average was changed by $\pm 10$\%, 
the resultant temperature stayed within $\pm 1$ keV ($\sim 30$\%
change).  In the case of changing the CXB level of the outer 
south-east region for investigating the directional dependence by 
$\pm 20$\%, the temperature also changed by $\pm 1$ keV 
($\sim 30$\% change). The normalizations of both the outermost regions 
for the azimuthal average and the south-east changed within 5\% and
15\%, respectively. The shapes of the ICM spectra around 2--3 keV 
and the CXB one with the power-law index of 1.4 are different from 
each other in the energy band. The ICM temperature in the outermost 
region, therefore, does not suffer significantly from the CXB level 
and fluctuation. As a result, even if we consider such uncertainties, 
the resultant values of the electron temperature, density, and the 
entropy for the ICM components do not significantly change by 
the NXB, CXB, and OBF contaminant systematics.

We also estimated the fraction of photons entering from outside of 
the extracted regions using a simulator of the Suzaku XRT/XIS system 
``xissim'' tool.  When a much brighter region is outside of the 
extracted region, photon contamination from the bright source would 
affect significantly in such cluster outskirts observations. Moreover, 
the point spread function of the Suzaku XRT has an extended tail 
(see e.g.,\ \cite{sato07}).  We, thus, need to estimate the fraction 
to confirm the results derived from the spectral fits. If we assumed 
a $\beta$-model of Abell 1246 emission, as described in section 
\ref{sec:data}, extending beyond the virial radius, the scattering 
and stray light contamination of the outermost region from other 
regions including the bright core would be within $\sim15$\% in the 
spectral fits.

\section{Results and Discussion}

\subsection{Temperature and Density Profiles}
\label{subsec:temp}

\begin{figure*}[thp]
\centerline{\FigureFile(0.7\textwidth,8cm){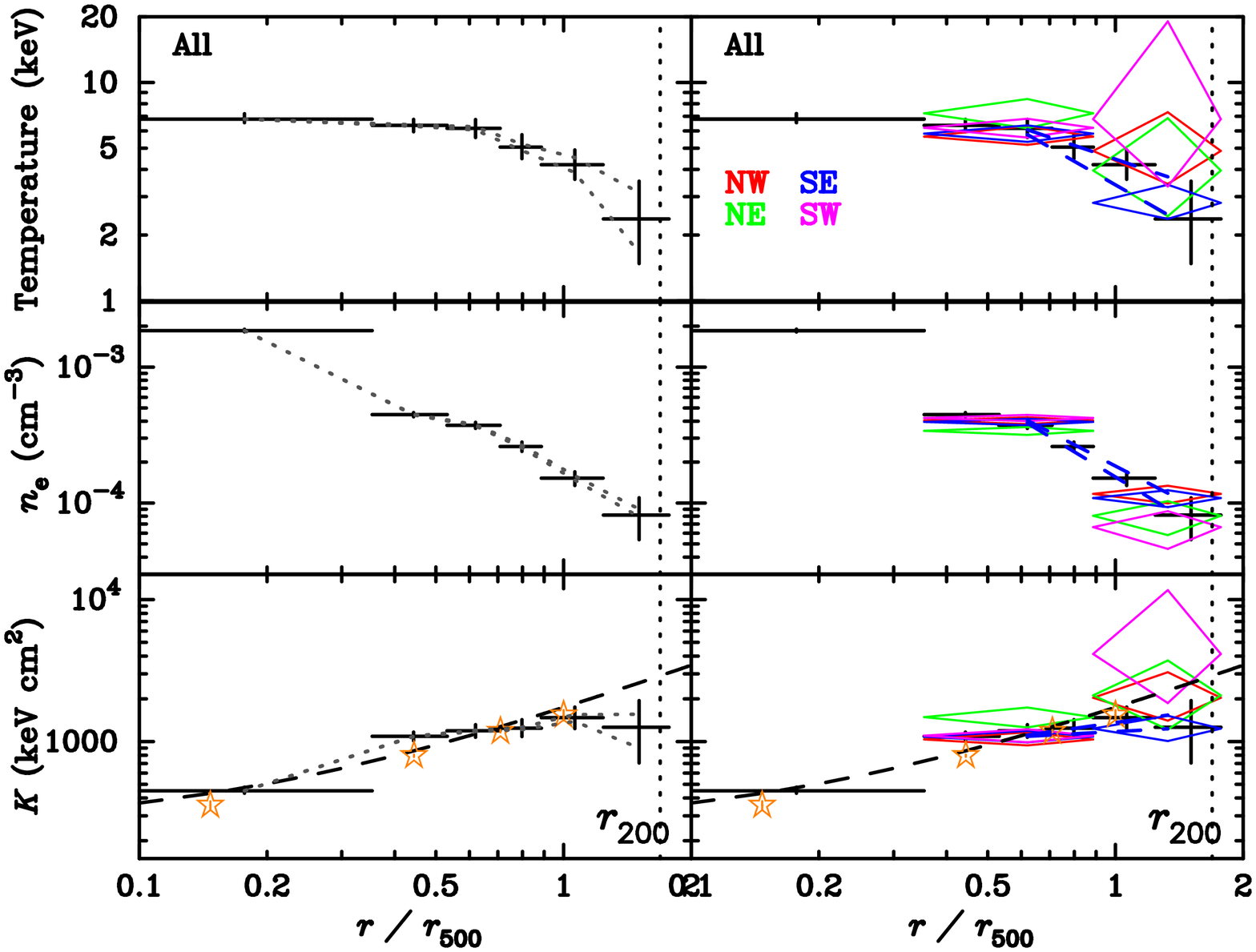}}
%\vspace*{-1ex}
\vspace*{3ex}
\caption{ 
Radial profiles of electron temperature (upper), density (middle), 
and entropy (lower) as a function of radius in units of $r_{500}$, 
where $r_{500}$ is defined as \timeform{6.1'} derived from the mass 
profile.  Black crosses show the values for the annular regions in 
each panel. Red, green, blue, and magenta diamonds in right panels 
correspond to the directional regions of northwest (NW), 
northeast (NE), southeast (SE), and southwest (SW) regions, 
respectively. The dashed lines in the lower panels indicate the best 
fit of the function $K \propto r^{1.1}$, as mentioned in subsection 
\ref{subsec:entropy}. Orange stars show the previous XMM-Newton 
results in \citet{pratt10}. Here, the dotted lines in the left panels 
and blue dashed lines in the right panels are corresponding
to the values by changing the CXB level by $\pm 10$\%, and $\pm 20$\% 
which is a comparable to the estimated CXB fluctuation, respectively.
}\label{fig:4}
\end{figure*}

The ICM temperatures in the annular regions of the cluster clearly 
decrease towards the virial radius.  The temperature in the region 
\timeform{7.56'}$<r<$\timeform{10.8'} (1.2--1.8~$r_{500}$) 
drops to $\sim$35\% of the peak temperature as shown in figure~\ref{fig:4}.  
The radial axis in figure~\ref{fig:4} is normalized by 
$r_{500} \sim$ \timeform{6.1'}, which is derived from the mass 
estimation of our observations under the H.E. assumption 
as mentioned in subsection \ref{subsec:mass}.
This decrease is slightly larger than the previous Suzaku results 
of other clusters
\citep{bautz09,reiprich09,hoshino10,sato10,akamatsu11}.  
The metal abundance is almost consistent at $\sim0.2$ solar from 
the central to the outer region of the cluster, although the 
abundance in $r>0.5~r_{500}$ has large errors.  The abundance of 
the central region is much lower than the values of the other clusters.  
These temperature and abundance are consistent with ASCA previous 
results, $5.17\pm0.58$ keV and $0.26\pm0.17$ solar in the whole 
cluster region, respectively, in \citet{fukazawa04}, but our results 
provides a better accuracy and radial distributions of the 
temperature and abundance. 

In order to investigate the directional dependence of the temperature, 
we derived the temperatures from the spectra of the four directions 
in the radius range of \timeform{2.16'}$<r<$\timeform{5.40'}
($0.4<r<0.9~r_{500}$) and \timeform{5.40'}$<r<$\timeform{10.8'} 
($0.9<r<1.8~r_{500}$) as shown in figure~\ref{fig:1} right.  
In the fits, the values of the CXB and Galactic components were fixed 
to those of the default case in table~\ref{tab:2}.  
The southeast direction tends slightly to have a lower temperature 
than the other directions, although those difference are within 
the statistical and systematic errors.  Note that the southeast 
direction has higher statistic because the region is covered by both 
the center and offset observations with Suzaku. 

We calculated the electron density from the normalization of 
the ICM spectral fits by considering the projection effect. 
The apec normalization parameter is defined as 
$Norm = 10^{14} \int n_{\rm e} n_{\rm H} dV / [4\pi (1+z)^2 D_{\rm A}^2]$
cm$^{-5}$, where $D_{\rm A}$ is the angular diameter distance to the 
source.  We estimated the deprojected $n_{\rm e} n_{\rm H}$ values 
assuming spherical symmetry and a constant temperature in each annular 
region and then assumed $n_{\rm e} = 1.2 n_{\rm H}$\@. We fitted 
the density profile with the $\beta$-model over all regions.  The 
derived $\beta$ value of the deprojected electron density profile is 
$\beta=0.47 \pm 0.02$, which agrees with the value in 
\citet{fukazawa04}.  The electron density of the annular 
regions to $r_{500}$ is also consistent with the previous XMM-Newton results 
for several clusters in $z<0.2$ \citep{croston08}.
We also investigated the directional difference of the electron 
density profile.  The deprojected electron densities of the northeast 
and southwest regions for $0.9<r<1.8~r_{500}$ tend to be lower than 
those of the southeast and northwest regions as shown in figure~\ref{fig:4}. 

\subsection{Mass Profile}
\label{subsec:mass}

\begin{figure*}[t]
\begin{center}
\begin{minipage}{0.43\textwidth}
\FigureFile(\textwidth,\textwidth){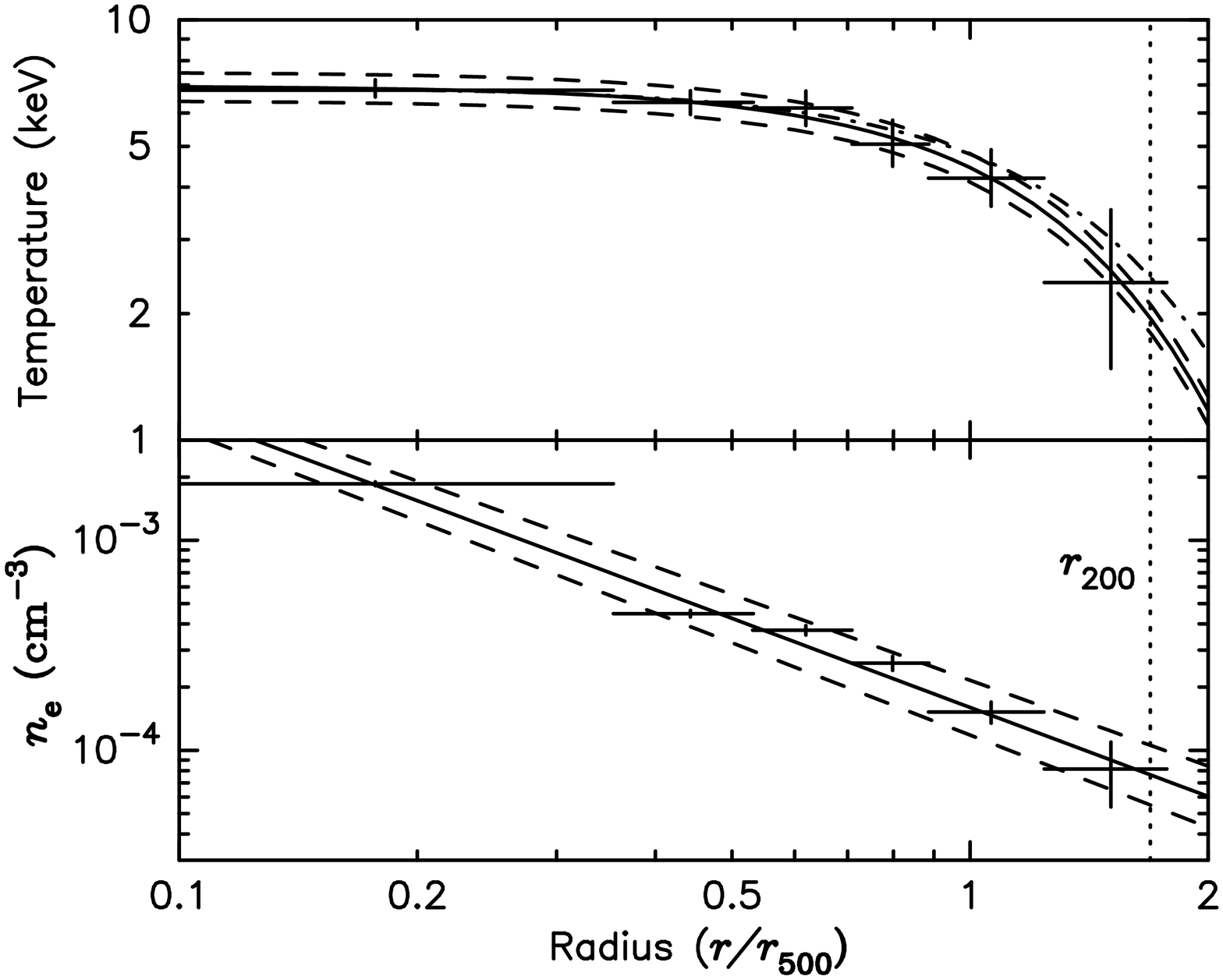}
\end{minipage}
\begin{minipage}{0.48\textwidth}
\FigureFile(\textwidth,\textwidth){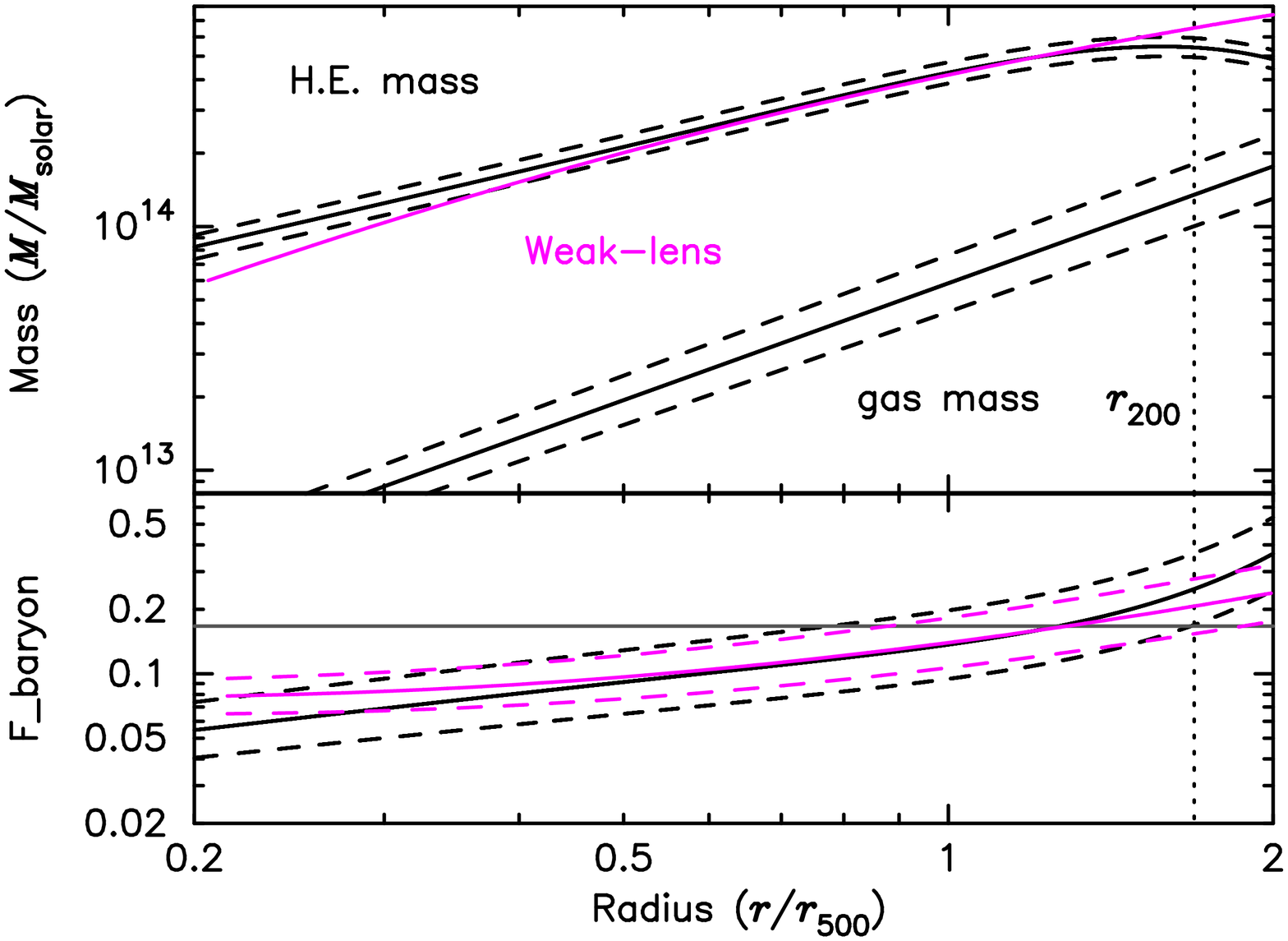}
\end{minipage}
%\vspace*{-1ex}
\vspace*{3ex}
\caption{
(Left) The resultant fits of the temperature (upper) and electron 
density (lower) as a function of radius in units of $r_{500}$. 
Solid and dashed lines correspond to the best fit and $\pm90$\% 
errors, respectively. The dash-dotted line shows the upper 
limit of the fits considering the CXB fluctuation by 10\%\@.
(Right) Upper panel: 
The H.E. mass and gas mass (upper and lower black solid lines, 
respectively) of Abell~1246 cluster from our data. Each dashed line 
shows $\pm90$\% errors. The magenta line shows the gravitational mass 
profile from weak-lens observations for lower mass cluster sample 
\citep{okabe10}.
Lower panel: 
Radial profile of the gas mass fraction to the H.E. mass (black) 
and the gravitational mass from weak-lens observations (magenta).
Dashed lines show $\pm90$\% errors.  The light gray line indicates 
the cosmic baryon fraction \citep{komatsu11}. 
The vertical dotted line corresponds to $r_{200}$ with the 
mean temperature by the formula in \citet{henry09}.
}\label{fig:5}
\end{center}
\end{figure*}

We derived the gravitational mass of Abell~1246 assuming spherical 
symmetry and H.E. (hereafter H.E. mass). Under the assumption, the 
total integrated gravitational mass $M_{<R}$ within the 
three-dimensional radius $R$ is given by 
\begin{eqnarray}
M_{< R} = - \frac{kTR}{\mu m_{\rm P} G} 
\left( \frac{d \ln \rho_{\rm gas}}{d \ln R} + \frac{d \ln T}{d \ln R} \right)
\end{eqnarray}
where $G$ is the gravitational constant, $\mu$ is the mean molecular
weight of the gas and $m_{\rm P}$ is the proton mass.  
We fitted the temperature and electron density profiles with 
the $\beta$-model formula separately as shown in figure \ref{fig:5}, 
and then derived the H.E. mass with the fitted parameters. 
The gas mass density $\rho_{\rm gas}$ is expected to be given as
$\rho_{\rm gas} = 1.92 \mu m_{\rm P} n_{\rm e}$
with $\mu=0.62$ and the ion density including 
helium is $n_{\rm i} = 0.92 n_{\rm e}$\@.  We also calculated 
the overdensity to the critical density of the universe 
from the derived H.E. mass and found the $r_{500} = $ \timeform{6.1'}.
On the other hand, the $r_{200} =$ \timeform{9.0'} derived from the
H.E. mass profile deviated by $\sim9$\% from those estimated $r_{200}$ 
by the empirical formula with the mean temperature in \citet{henry09}.

The radial mass profile of figure~\ref{fig:5} is normalized by 
the radius of $r_{500}$ to enable easy comparison with the previous 
results.  The errors of the total gas mass profiles are derived 
from the sum of 90\% errors of the fitted parameters of the temperature 
and electron density with the $\beta$-model formulae.  Note that the 
resultant H.E. mass starts flattening or decreasing beyond the $r_{500}$ 
region.  One plausible cause is the systematic effect from the too
simple model formula, such as a simple $\beta$-model. However, the 
other formulae also result in the similar feature for other clusters.
The observed steep temperature drop is causing such a mass distribution
(see also \cite{bonamente13, ichikawa13}).  Consequently, this feature 
would indicate a flaw in the H.E. assumption in $r>r_{500}$.
Even if we use the temperature profile including the uncertainties of 
the CXB level by 10\%, the feature does not change as shown by the 
dash--dotted line in figure \ref{fig:5} left. As mentioned in subsection 
\ref{subsec:entropy}, the flatness or decrease of the entropy in 
$r>r_{500}$ would also indicate being out of H.E. in the outskirts 
region of the cluster. Again, we note that although the derived mass 
indicates the azimuthal averaged mass, it would be affected by the 
higher statistic of the southeast direction.

The derived H.E. mass within $r<r_{500}$ from our data is 
$(4.3 \pm 0.4) \times 10^{14}~M_{\odot}$. It is consistent with 
that by \citet{vikhlinin09}, 
$(3.9 \pm 0.1) \times 10^{14}~M_{\odot}$ at $r_{500}$ 
derived with the $M_{500}$--$T_{\rm X}$ scaling relation
with Chandra observations.  It is useful to compare 
the cluster mass with several methods to avoid systematic bias.
We, therefore, compare the H.E. mass with the gravitational 
mass from weak-lens observation.  \citet{okabe10} derived the 
gravitational mass profile based on the Navarro-Frenk-White (NFW) 
model \citep{navarro96} for the lower and higher mass cluster sample 
with $\sim 10$\% relative accuracies. When we compare the H.E. mass 
profile from our data with the mass profiles from the lower and 
higher mass sample, the former profile agrees well. Using the 
parameters for the lower mass sample, the calculated gravitational 
mass within $r_{500}$ and $r_{200}$ are 4.2 and 
6.5$\times 10^{14}~M_{\odot}$, respectively.
As shown in figure \ref{fig:5}, the H.E. mass and the mass from the 
lower mass sample are fairly consistent within $r_{500}$, and beyond 
this radius the mass from weak-lens is larger than that from our data.

The fraction of the derived gas mass to the H.E. mass from our data 
around $r_{500}$ is $14^{+6}_{-5}$\%, which is consistent with 
the cosmic baryon fraction value of $16.7$\% \citep{komatsu11} within 
the statistical error. As described above, because the $r>r_{500}$ 
region is apparently not in H.E., we calculated the fraction at 
$r_{200}$ with the gravitational mass from weak-lens observation. 
The gas mass at $r_{200}$, is derived to be
$(1.4 \pm 0.4) \times 10^{14}~M_{\odot}$ from our observation, and 
the gas mass fraction using the weak-lens template model is 
$(21 \pm 5)$\%, which also agrees with the cosmic baryon fraction. 
Because the typical scattering of the relation between 
the cluster mass and the concentration parameter of the NFW model
for lower mass cluster sample in \citet{okabe10} is $\sim 10$\%,
the fraction still agrees with the cosmic baryon fraction.

In the region around $r_{500}$, we compared the slope of the mass 
density of Abell~1246 with the expected slope $\rho \propto r^{-3}$ 
from the NFW profile. The mass density slope from our result seems 
to be steeper than the one derived from the NFW model.  
\citet{kawaharada10} and \citet{akamatsu11} also suggest 
such a steeper mass profile in the outer region of the clusters.
This will be real but systematic error such as H.E. assumption 
validity shall be critically reviewed, which is beyond the scope 
of this paper. If the H.E. assumption was invalid in this region,
we would have to estimate the cluster mass, particularly in the 
outskirts, with weak-lens observations.

\subsection{Entropy Profile}
\label{subsec:entropy}

An entropy profile provides the thermal process and history of the 
ICM, particularly for the gas heated by the accretion shock from 
outside of the cluster.  In X-ray astronomy, we define the entropy 
as $K=kTn_{\rm e}^{-2/3}$\@.  The resultant entropy profile in the 
annular regions is shown in the lower panels of figure~\ref{fig:4}.  
The entropy increases with radius to $\sim r_{500}$, and the profile 
has a flatter slope at $r>r_{500}$\@.  This tendency is consistent 
with previous Suzaku results 
\citep{bautz09,george09,hoshino10,akamatsu11}.   
Compared to the previous XMM-Newton results of 31 clusters within 
$r_{500}$ in \citet{pratt10}, our results agree with the entropy 
profile within $r_{500}$ as shown in figure~\ref{fig:4}.  We also 
investigated the directional difference of the entropy in the same 
manner as the temperature and electron density profiles.  Although 
the values in the outermost region have large errors, the entropy 
of the southeast direction tends to be lower than those of the other 
directions.

\begin{figure*}[t]
\begin{minipage}{0.45\textwidth}
\FigureFile(\textwidth,\textwidth){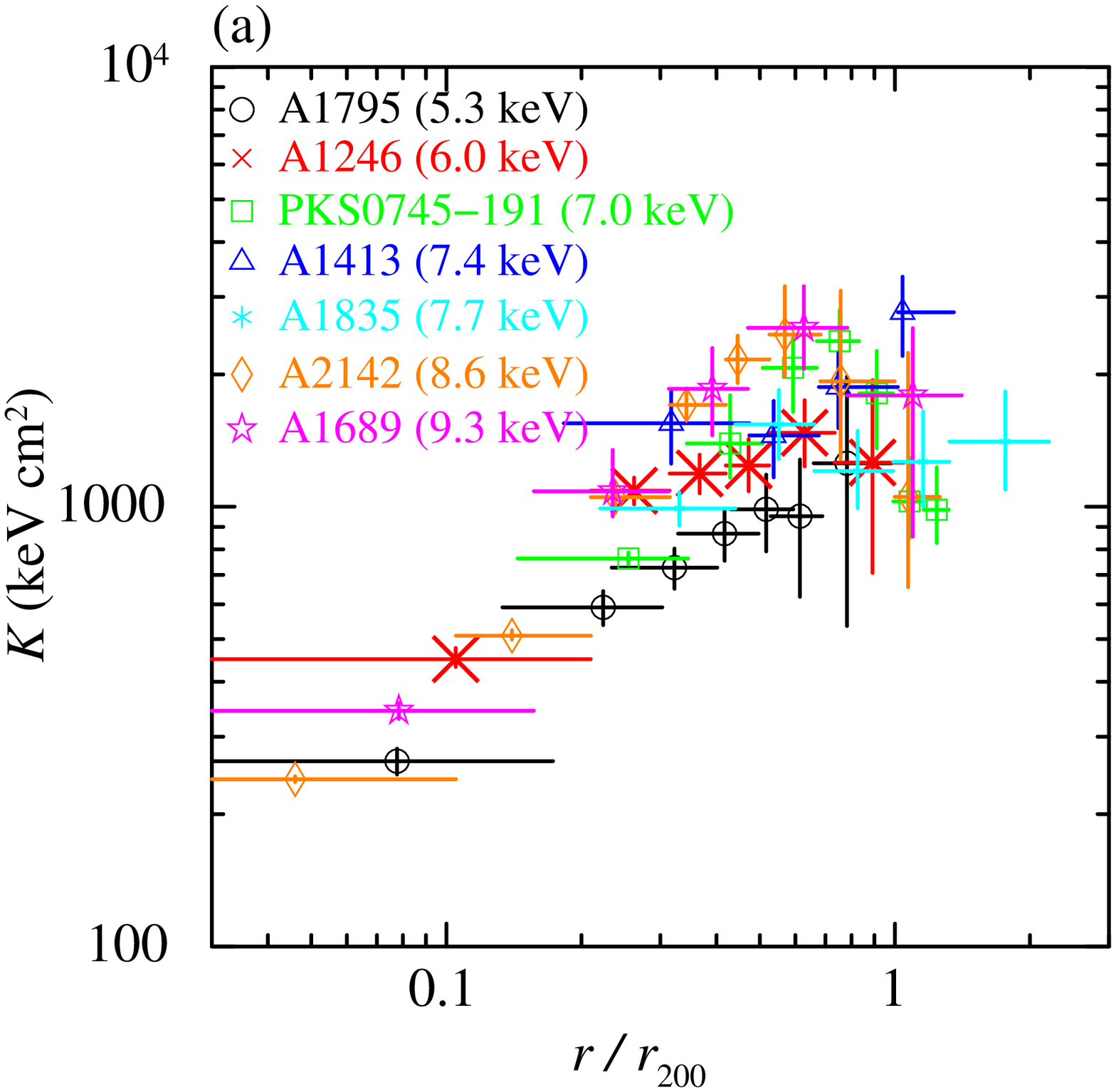}
\end{minipage}\hfill
\begin{minipage}{0.47\textwidth}
\FigureFile(\textwidth,\textwidth){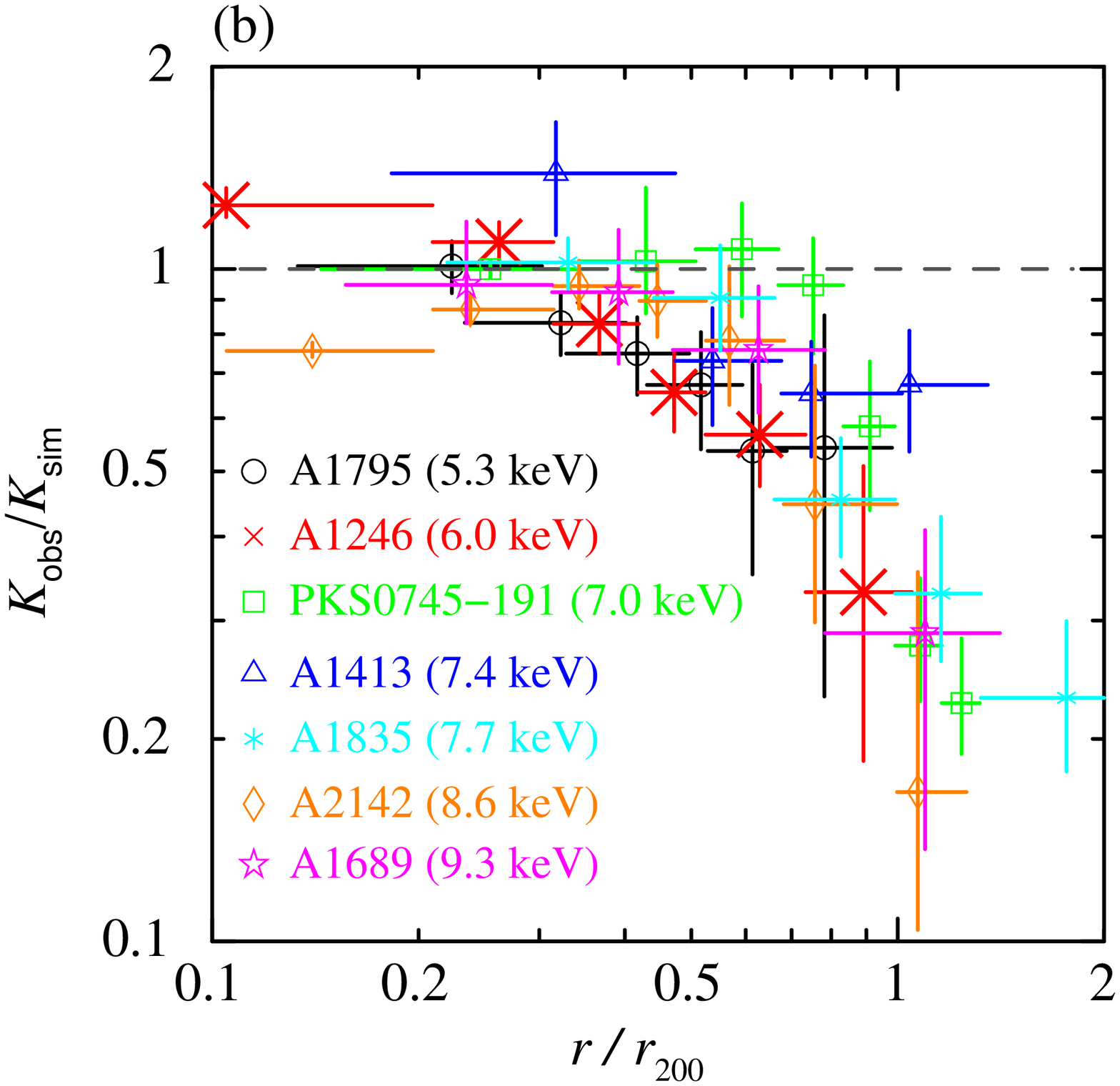}
\end{minipage}
%\vspace*{-1ex}
\vspace*{3ex}
\caption{(a) Derived radial entropy profile for each cluster from 
Suzaku observations. The radii are normalized by $r_{200}$ with the 
mean temperature as shown in \citet{henry09}.
(b) Ratios of the entropy from X-ray observations to the expected 
values from the simulations. Dashed line shows the entropy from the 
simulations.
}\label{fig:6}
\end{figure*}

\citet{voit05} reports $K \propto r^{1.1}$ on the basis of numerical 
simulations of adiabatic cool gas accretion, and the XMM-Newton results 
\citep{pratt10} agree with the relation within $r_{500}$. 
We, therefore, fitted our entropy profile for the annulus regions 
by a power-law model with an index of 1.1\@.  The best fit is shown 
by the dashed line in the lower panel of figure~\ref{fig:4}.  
Our result matches well with the model within $r_{500}$, while, 
in $r>r_{500}$, our result tends to have a smaller index.  
As for directional dependence, the entropy of the southeast direction 
shows a smaller index than the model, and the indexes of the other 
directions agree with the model, although they have large errors. 

We compared our result with other results from Suzaku, 
Abell~1795 \citep{bautz09}, PKS0745-191 
\citep{george09, walker12}, Abell~1413 \citep{hoshino10}, 
Abell~1835 \citep{ichikawa13}, Abell~2142 \citep{akamatsu11}, 
Abell~1689 \citep{kawaharada10}, 
as shown in figure \ref{fig:6} (a).  
Here, the radius is normalized by $r_{200}$ with the 
mean temperature as shown in figure \ref{fig:6}, 
for each cluster, using the equation in \citet{henry09}.
All the clusters observed with Suzaku have a similar tendency that 
the entropy increases with radius until $r_{500} \sim 0.5~r_{200}$, 
and the profile has a flatter slope in $r>r_{500}$.
In order to correct the subtle mass dependence of the profile, we 
examined ratios of the derived entropy from Suzaku observations to 
the expected values from the numerical simulation in \citet{voit05}, 
for all the clusters as shown in figure \ref{fig:6} (b).
As a result, all the clusters regardless of the system size have 
a similar deviation tendency in $r>0.5~r_{200}$.

Plausible causes of the flattening of the entropy profile, which are 
explained in previous Suzaku papers, are a flaw in the H.E. assumption, 
clumpiness, or both in the outer region of clusters.  
\citet{kawaharada10} suggest that the kinetic motions such as bulk 
or turbulence motions are required under the condition.
As mentioned in \citet{simionescu11}, the clumpiness which comes from 
the accreting gas from the filamentary structure in the universe could 
overestimate the gas density \citep{nagai11} and increase the 
flattening of the entropy profile in the outer region of 
clusters.  In Abell~1246 cluster, however, entropy flattening 
appears in the southeast region rather than the northwest region, which 
appears to be accreting from the filament.  Another plausible cause is 
a difference between ion and electron temperatures in the outer
low-density region of clusters, while the electron temperature is 
equal to the ion temperature in the central region of the clusters, 
because the equilibration timescale for electron-ion collisions is 
much longer than the elapsed time after the shock heating
\citep{hoshino10, akamatsu11}. 

In near future, X-ray microcalorimeter, such as the SXS instrument 
on ASTRO-H \citep{mitsuda10}, which has a 20--30 times higher energy 
resolution than CCD instruments, allows us to investigate the kinetic 
motions and the difference between the electron and ion temperatures. 
However, because of the small effective area of the SXS, these values 
are accessible only in the core region of the bright cluster such as 
the Perseus cluster.  For the observations of the faint region 
such as the cluster outskirts, we would need to wait for the 
satellite with the larger effective area and field of view, such 
as DIOS \citep{ohashi10}. For resolving the clumpiness, the imaging 
analysis with the higher angular resolution would be needed, or 
the comparison of the density from the X-ray and 
Sunyaev-Zel'dovich effect (e.g.,\ \cite{sunyaev70,sunyaev72}) 
observations could be also useful.

\subsection{Upper Limit of Oxygen Emission Lines}
\label{subsec:upperlimit}

\begin{figure}[tb]
\centerline{\FigureFile(0.5\textwidth,8cm){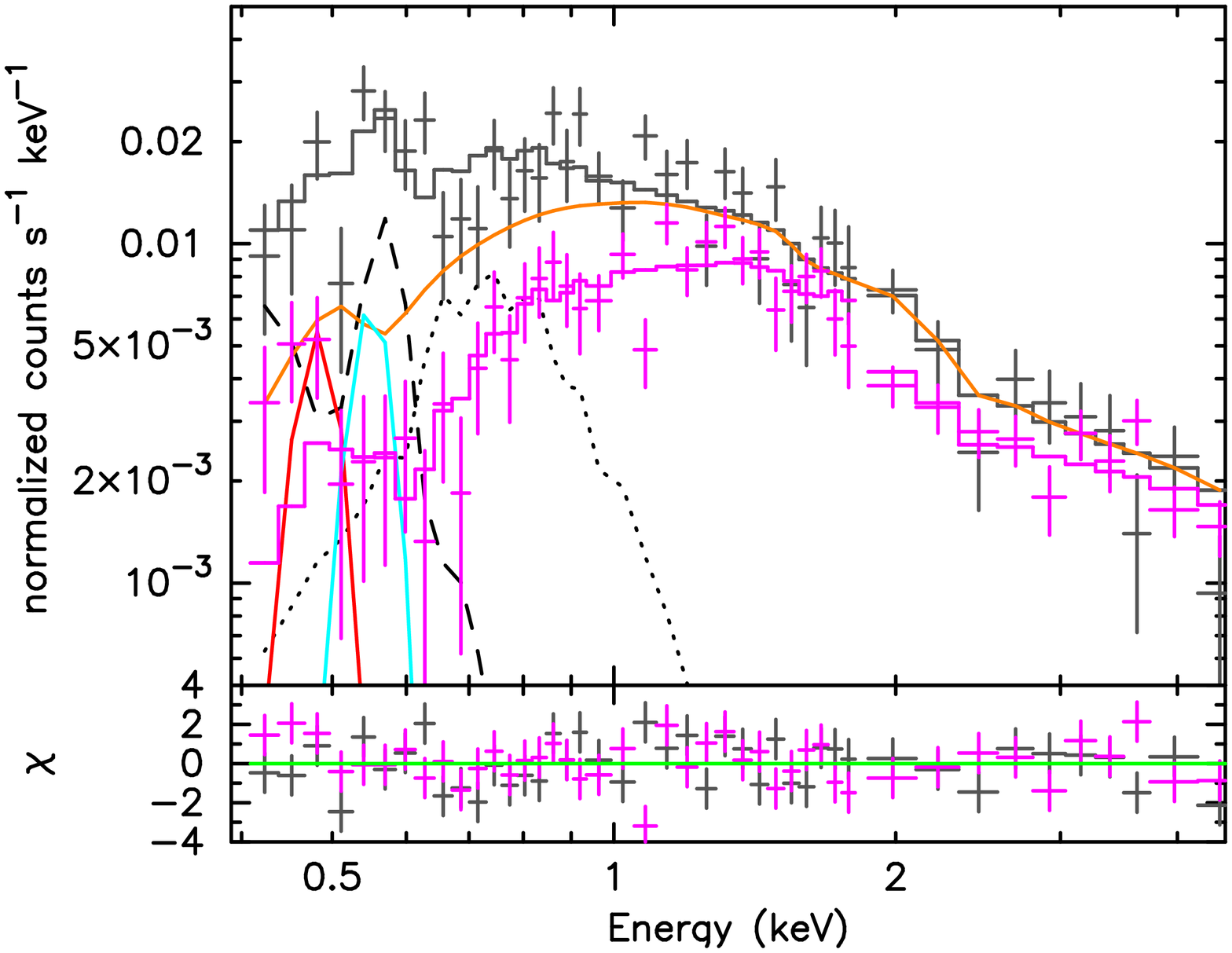}}
%\vspace*{-1ex}
\vspace*{5ex}
\caption{Panel showing the resultant fits for the constraint of the 
intensities of the \emissiontype{OVII} and \emissiontype{OVIII} lines 
in the outside region of the virial radius. The \emissiontype{OVII} 
and \emissiontype{OVIII} emission lines are shown by red and 
light-blue lines, respectively, and the notations 
of the other lines are the same as in figure \ref{fig:3}.
}
\label{fig:7}
\end{figure}

\begin{table}
\caption{Resultant intensities of O\emissiontype{VII} and
O\emissiontype{VIII} emission lines with a 2$\sigma$ confidence range.}
\label{tab:4}
\begin{center}
\begin{tabular}{lc} \hline 
O\emissiontype{VII} & \\
Center (keV) at $z=0.1902$& 0.482(fixed) \\
Sigma & 0 (fixed) \\
I ($\times10^{-7}$ photons cm$^{-2}$ s$^{-1}$ arcmin$^{-2}$) & 1.3$^{+1.6}_{-1.3
}$ \\
\hline
O\emissiontype{VIII} & \\
Center (keV) at $z=0.1902$ & 0.549(fixed) \\
Sigma & 0 (fixed) \\
I ($\times10^{-7}$ photons cm$^{-2}$ s$^{-1}$ arcmin$^{-2}$) & 2.1$^{+3.5}_{-2.1
}$ \\
\hline\\
\end{tabular}
\end{center}
\end{table}

The warm gas in the large-scale filament is an important part of 
the WHIM which is the dominant component of baryons in the local 
universe (e.g.,\ \cite{yoshikawa03}).  Although Suzaku observations 
in several clusters of galaxies and superclusters have been 
performed to search for the redshifted O emission lines from the WHIM, 
no positive detection has been obtained. We fitted the spectra taken 
outside of the virial radius, $r>$\timeform{10.8'}~$\sim r_{200}$
of the cluster.  This region is the same as the one used to estimate 
the fore- and background emissions, as shown in 
subsection \ref{subsec:eachspectra}. 
Although the spectra in this region were well represented by the Galactic 
and CXB components without an additional model as mentioned in subsection 
\ref{subsec:eachspectra}, we constrained the upper limit of 
the electron density of the thermal model of $\sim 1$ keV (model III), 
$7.5\times10^{-5}$ cm$^{-3}$ as the 2$\sigma$ confidence limit, 
under the assumption that the ICM emission extended to the radius with 
a spherical symmetry. If the gas emission came from the filamentary 
structure, i.e. WHIM, in the line of sight depth of 20 Mpc, 
the $n_{\rm H}$ would be 
$8.5 \times 10^{-5}~{\rm cm}^{-3}~(L/20~{\rm Mpc})^{-1/2}$.
Assuming that the gas temperature from the filamentary structure 
should be much lower, we investigated the upper limit 
of the intensities of the additional O\emissiontype{VII} and 
O\emissiontype{VIII} lines at the cluster redshift from the WHIM 
origin, because those emission lines would be more sensitive than 
the thermal component to constrain the WHIM signal in the lower 
temperature. 
We fitted the data with the following model: 
${\rm apec}_{\rm LHB}+{\rm phabs}\times({\rm apec}_{\rm MWH}+
{\rm pow}_{\rm CXB}+{\rm gaussian}_{\rm OVII}+{\rm gaussian}_{\rm OVIII})$\@.
Here, the parameters of the Galactic and CXB components followed 
our previous fit values in the default case of table~\ref{tab:2}.  
The temperatures of the Galactic components and the photon index 
of the CXB component were fixed, and the normalizations were free 
in the fit. The redshifted central energies of the O\emissiontype{VII} 
and O\emissiontype{VIII} Gaussian lines were fixed at 482 and 549 eV, 
respectively, with no intrinsic width of the lines assumed.  We 
employed an increment of $\delta \chi^2 =4$ as the measure for the 
2$\sigma$ upper limits of the line intensities.  This value also 
corresponds to the 95\% limit for an additional single parameter 
(single line intensity) in the F-test.  As a result, we determine 
the upper limit of the O\emissiontype{VII} and O\emissiontype{VIII} 
lines to be 2.9 and 5.6 $\times 10^{-7}$ photons cm$^{-2}$ s$^{-1}$ 
arcmin$^{-2}$ as the 2$\sigma$ confidence limits as shown in 
table~\ref{tab:4} and figure~\ref{fig:7}.
The resultant values are consistent with the results
in \citet{takei07}, \citet{sato10}, and \citet{mitsuishi12} 
within 2$\sigma$ error.  Assuming 20 Mpc for the line-of-sight 
depth of the WHIM distribution, we constrain the density of the
WHIM cloud under the condition of the temperature to be 
$T=2\times10^6$ K. Using the line intensity, $I<2.9\times 10^{-7}$ 
photons cm$^{-2}$ s$^{-1}$ arcmin$^{-2}$ at $z=0.1902$, and the 
ratio of electron to hydrogen number density of 
$n_{\rm e}/n_{\rm H}=1.2$ for ionized gas the following condition
is obtained:
\begin{eqnarray}
n_{\rm H} &<& 4.7\times 10^{-5}~{\rm cm}^{-3}~(Z/0.2~Z_{\odot})^{-1/2}~
(L/20~{\rm Mpc})^{-1/2}.
\label{eq:1}
\end{eqnarray}
The corresponding overdensity $\delta \equiv n_{\rm H}/\bar{n}_{\rm H}$ is
\begin{eqnarray}
\delta&<&160~ (Z/0.2~Z_{\odot})^{-1/2}~(L/20~{\rm Mpc})^{-1/2}.
\label{eq:2}
\end{eqnarray}
Our upper limits for the O\emissiontype{VII} and
O\emissiontype{VIII} lines are consistent with the previous 
Suzaku observations \citep{takei07, tamura08}. 
As for the WHIM search, a new mission which can 
separate the WHIM originated OVII and OVIII emission lines 
(e.g. the DIOS mission, \cite{ohashi10}) will be needed.

\subsection{Interpretation of the Morphology of Abell~1246}

The X-ray image of Abell 1246 is obviously elongated to the 
northwest-southeast direction. This follows the galaxy density 
map, whose ratio of the major to minor axis of the density is about 2, 
as shown in figure~\ref{fig:1}.  If the intracluster gas is under 
the condition of H.E. in the gravitational potential dominated by the 
dark matter, gas distribution should deviate from the spherical symmetry.  

Comparing to Abell 1689 reported in \citet{kawaharada10}, the 
temperatures for Abell~1246 and Abell~1689 in the direction which 
connects to an overdense filamentary structure of galaxies, are 
higher than those in the other direction. However, the temperatures 
in the other direction for Abell~1246 are relatively high unlike 
with those in Abell~1689.  As shown in figure~\ref{fig:1} right, 
the galaxy distribution around Abell~1246 is extended to
the northwest--southeast direction.  In fact, a comparison of the 
northwest and southeast regions reveals that the temperature and 
entropy of the northwest region are higher than those of the southeast, 
while both electron densities are consistent. While the entropy of 
the southeast region has a flatter slope, the entropy of the northwest 
region agrees with slope from the numerical simulation.  

If the gas infalls in an asymmetric manner from the filament onto 
the relaxed cluster and the accreted gas is not mixed with the 
original intracluster gas, the gas profile deviates from the 
spherical symmetry and the X-ray morphology becomes elliptical.  
In fact, \citet{kawaharada10} find a hot spot in the direction 
of the filament, and our results suggest the same condition. 
These facts would imply the same scenario of the accreting matter 
to the clusters.

\section{Summary}

We studied the electron temperature, density, cluster mass and entropy 
profiles in Abell~1246 cluster and around the cluster outskirts beyond 
the virial radius observed with Suzaku.  We summarize the resultant 
features of Abell~1246 cluster as follows;
\begin{itemize}
\item The temperature drops from $\sim$7 keV at the central region 
to $\sim$2.5 keV around $r_{200}$ region of the cluster.

\item The calculated total mass within $r_{500}$ under the H.E. 
assumption is $(4.3 \pm 0.4) \times 10^{14}~M_{\odot}$ 
and the gas mass fraction agrees with the cosmic baryon fraction.

\item The derived entropy profile has a flatter slope compared to the 
expected slope from the numerical simulation in $r>r_{500}$. 

\item 
In order to compare our results with other cluster results with Suzaku,
we investigated the ratios of the observed entropy to the expected value 
from the numerical simulation. The resultant radial entropy ratios for each 
cluster have a similar tendency. 

\item We constrain the intensities of O\emissiontype{VII} and 
O\emissiontype{VIII} lines at the cluster redshift to be less than 
2.9 and 5.6 $\times 10^{-7}$ photons cm$^{-2}$ s$^{-1}$ 
arcmin$^{-2}$, respectively, as 2$\sigma$ upper limits.  The 
intensity of O\emissiontype{VII} indicates 
$n_{\rm H}< 4.7\times 10^{-5}$ cm$^{-3}$ ($Z/0.2~Z_{\odot}$)$^{-1/2}$
($L/20~{\rm Mpc}$)$^{-1/2}$, 
which corresponds to the overdensity,
$\delta<160$ ($Z/0.2~Z_{\odot}$)$^{-1/2}$ ($L/20~{\rm Mpc}$)$^{-1/2}$\@.
\end{itemize}

In the near future, X-ray microcalorimeter missions such as DIOS would 
give a lot of hints for investigating cluster outskirts. Also, we would 
have to estimate the cluster properties in several ways 
without bias effects. 

\bigskip
We thank the referee for providing valuable comments and suggestions.
We acknowledge the support by a Grant-in-Aid for Scientific Research 
from the MEXT, No.25800112 (K.S).

\appendix

\section{Comparisons of the CXB intensity with the previous results}
\label{app:1}

We estimate the CXB surface brightness in our observations to be 
5.88$^{+0.41}_{-0.38} \times 10^{-8}$ erg cm$^{-2}$ s$^{-1}$ sr$^{-1}$ 
from the spectral fit after the point-like sources subtractions 
as shown in table \ref{tab:2}. The previous ASCA result 
\citep{kushino02} shows the CXB surface brightness,
$(6.38 \pm 0.07 \pm 1.05) \times 10^{-8}$ 
erg cm$^{-2}$ s$^{-1}$ sr$^{-1}$, 
(90\% statistical and systematic errors) 
with the photon index, 1.412 in 2--10 keV\@.
\citet{moretti09} also summarize the CXB level, including their new 
result with SWIFT. The derived CXB level in 2--10 keV from SWIFT
is $(7.16 \pm 0.43) \times 10^{-8}$ 
erg cm$^{-2}$ s$^{-1}$ sr$^{-1}$ with the photon index, 
$\Gamma=1.47 \pm 0.07$.
The measured CXB surface brightnesses show a significant 
range from the HEAO-1 value of (5.41$\pm$0.56)$\times10^{-8}$ 
erg cm$^{-2}$ s$^{-1}$ sr$^{-1}$ \citep{gruber99}
to (7.71$\pm$0.33)$\times 10^{-8}$ erg cm$^{-2}$ s$^{-1}$ sr$^{-1}$ 
with SAX-MECS \citep{vecchi99} in 2--10 keV\@. 
Because these measurements show the surface brightness to be 
within about 10\% of the level reported by \citet{kushino02} with ASCA, 
we compare our estimation primary with those of 
\citet{kushino02}, 
$I_0 = (6.38 \pm 0.07 \pm 1.05) \times 10^{-8}$ 
erg cm$^{-2}$ s$^{-1}$ sr$^{-1}$. 
Here, we calculated the integrated point source flux per steradian from
\begin{eqnarray}
\label{eq:fs}
I_{ps} (S>S_{0}) &=& \frac{k_0}{\gamma -2}\; S_{0}^{-\gamma+2},
\end{eqnarray}
where $k_0$ and $\gamma$ are the differential $\log N$--$\log S$ 
normalization and slope, respectively. We took nominal values,
$k_0=1.58\times 10^{-15}$~sr$^{-1}$~(erg cm$^{-2}$ s$^{-1}$)$^{\gamma-1}$
and $\gamma=2.5$, from \citet{kushino02} as shown in \citet{hoshino10}.
$S_{0}$ was taken as $3\times 10^{-14}$ erg cm$^{-2}$ s$^{-1}$,
which corresponds to the faintest flux level of the point-like source 
in our analysis as mentioned in subsection~\ref{subsec:pointsource}. 
Note that the assumed $\log N$--$\log S$ in equation (\ref{eq:fs}) does 
not take into account the flattening of the relation in the fainter 
flux end. The expected CXB surface brightness is 
$I_0 - I_{ps} = 4.56 \times 10^{-8}$ erg cm$^{-2}$ s$^{-1}$ sr$^{-1}$.

In addition, to estimate the amplitude of the CXB 
fluctuations, we also scaled the measured fluctuations from Ginga 
\citep{hayashida89} to our flux limit and the field of view (FOV) area. 
The fluctuation width is given by the following relation,
\begin{eqnarray}
\frac{\sigma_{\rm Suzaku}}{I_{\rm CXB}} = \frac{\sigma_{\rm Ginga}} 
{I_{\rm CXB}} \left(\frac{\Omega_{\rm e,Suzaku}}{\Omega_{\rm e,Ginga}} 
\right)^{-0.5} \left(\frac{S_{\rm c,Suzaku}}{S_{\rm c,Ginga}} 
\right)^{0.25},
\end{eqnarray}
where $(\sigma_{\rm Suzaku}/I_{\rm CXB})$ means the fractional CXB
fluctuation width due to the statistical fluctuation of discrete
source number in the FOV\@. Here, we adopted
$\sigma_{\rm Ginga}/I_{\rm CXB} = 5\%$,
with $S_{\rm c}$ (Ginga: $6\times10^{-12}$ erg cm$^{-2}$ s$^{-1}$)
representing the upper cut-off of the source flux,
and $\Omega_{\rm e}$ (Ginga: 1.2~deg$^2$) representing
the effective beam size (or effective solid angle) of the detector.
The derived $\sigma_{\rm Suzaku}$/$I_{\rm CXB}$ was 4.9\% with 
$\Omega_{\rm e,Suzaku}=0.09$ deg$^2$ for the Suzaku FOV, 
and $S_{\rm c,Suzaku} = 3 \times 10^{-14}$ erg cm$^{-2}$ s$^{-1}$.
As for the background region ($r>$\timeform{10.8'}) and the 
cluster outermost region (\timeform{7.56'}$<r<$\timeform{10.8'}) 
of the spectral fits for the azimuthal average,
$\Omega_{\rm e} = 0.04$ and 0.03 deg$^2$,
we examined the fluctuation level to be
12.0\% and 13.9\% in the 90\% confidence region, respectively.
We also estimated the fluctuation from HEAO-1 A2 results 
\citep{shafer83} with $\sigma_{\rm HEAO-1}/I_{\rm CXB} = 2.8\%$,
$\Omega_{\rm e,HEAO-1}=15.8$ deg$^2$, and 
$S_{\rm c,HEAO-1} = 8 \times 10^{-11}$  erg cm$^{-2}$ s$^{-1}$.
The derived $\sigma_{\rm Suzaku}$/$I_{\rm CXB}$ for the 
background region with the HEAO-1 results was 17.3\%. 
This value was slightly 
larger than that with ASCA. As for the directional dependence, 
because the CXB fluctuation of the eastern outer region, 
\timeform{5.40'}$<r<$\timeform{10.8'} was 17.0\% with 
$\Omega_{\rm e,Suzaku}=0.02$ deg$^2$, 
we estimated the uncertainties from the fluctuation changing the 
CXB level by $\pm 20$\% from the Ginga results. 
The resultant uncertainties are shown in figure \ref{fig:4}.

As a result, the best-fit parameter of the CXB 
surface brightness for the background region, $r>$\timeform{10.8'} 
(after subtraction of point sources brighter 
than $3\times10^{-14}$ erg cm$^{-2}$ s$^{-1}$ in 2--10 keV band) 
is $(5.88^{+0.41}_{-0.38} \pm 0.7) \times 10^{-8}$ erg cm$^{-2}$ s$^{-1}$ 
sr$^{-1}$ which agrees with those of the previous Suzaku results 
\citep{hoshino10, ichikawa13}, although our resultant CXB surface 
brightness is slightly larger than the value expected one from 
the ASCA results ($4.56 \times 10^{-8}$ erg cm$^{-2}$ s$^{-1}$
sr$^{-1}$) with 
the 90\% statistical errors taking into account the CXB fluctuation.
One plausible cause would be the contaminations from the 
excluded point-like source signals and unresolved sources.

\end{document}